\documentclass[a4paper,fleqn,usenatbib]{mnras}

\usepackage{newtxtext,newtxmath}

\usepackage[T1]{fontenc}
\usepackage{ae,aecompl}

\usepackage{graphicx}	
\usepackage{amsmath}	
\usepackage{nccmath}
\usepackage{amssymb}	
\usepackage{nicefrac}
\usepackage{xspace}
\usepackage[super]{nth}
\usepackage[acronym]{glossaries}
\usepackage{array}
\usepackage[math]{cellspace}   
\usepackage[normalem]{ulem}

\graphicspath{{figures/}}

\newcommand{\sourceNameLong}{MASTER~OT~J061451.70--272535.5\xspace}
\newcommand{\sourceName}{J0614--27\xspace}

\newcommand{\python}{{\sc python}\xspace}
\newcommand{\pySHOC}{{\sc pySHOC}\xspace}

\newcommand{\HeII}{\ion{He}{II}\xspace}
\newcommand{\HeI}{\ion{He}{I}\xspace}

\renewcommand{\deg}{
  {\ensuremath{\,^{\circ}}}
}
\newcommand{\dms}[3]{
  \ensuremath{{#1}\deg{#2}^{\mathrm{'}}{#3}^{\mathrm{''}}}
}
\newcommand{\hms}[3]{
  \ensuremath{{#1}^{\mathrm{h}}{#2}^{\mathrm{m}}{#3}^{\mathrm{s}}}
}
\newcommand{\uNrTbl}[2]{
  #1&#2
}

\newcommand{\Mo}{\ensuremath{\, M_{\odot}}}
\newcommand{\Ro}{\ensuremath{\, R_{\odot}}}

\newcommand{\orcid}[1]{%
	\href{https://orcid.org/#1}{\includegraphics[scale=0.2]{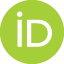}}%
    }

\newcommand{\ie}{i.e.\@ }
\newcommand{\eg}{e.g.\@ }

\newcommand{\wrt}{w.r.t.\@ }

\newacronym{SAAO}{SAAO}{South African Astronomical Observatory}
\newcommand{\SAAO}{\gls*{SAAO}\xspace}
\newacronym{SALT}{SALT}{South African Large Telescope}
\newcommand{\SALT}{\gls*{SALT}\xspace}
\newacronym{SHOC}{SHOC}{Sutherland High-speed Optical Camera}
\newcommand{\SHOC}{\gls*{SHOC}\xspace}
\newacronym{RSS}{RSS}{Robert Stobie Spectrograph}
\newacronym{CRTS}{CRTS}{Catalina Real-time Transient Survey}
\newacronym{MASTER}{MASTER}{Mobile Astronomical System of Telescope-Robots}
\newcommand{\MASTER}{\gls*{MASTER}\xspace}

\newacronym{CV}{CV}{cataclysmic variable}
\newacronym{mCV}{mCV}{magnetic cataclysmic variable}
\newcommand{\mCV}{\gls*{mCV}\xspace}
\newcommand{\mCVs}{\glspl*{mCV}\xspace}
\newcommand{\MCVs}{\Glspl*{mCV}\xspace}
\newacronym{IP}{IP}{Intermediate polar}
\newcommand{\IPs}{\glspl*{IP}\xspace}
\newacronym{WD}{WD}{white dwarf}
\newcommand{\WD}{\gls*{WD}\xspace}

\newacronym{VWF}{VWF}{Very Wide Field}
\newacronym{FWHM}{FWHM}{full width at half-maximum}
\newacronym{PSF}{PSF}{point source function}

\newacronym{SOM}{SOM}{self-organizing map}
\newacronym{QPOs}{QPOs}{Quasi-periodic oscillations}
\newcommand{\qpos}{\gls*{QPOs}\xspace}
\newacronym{AML}{AML}{angular momentum loss}

\newacronym{MB}{MB}{magnetic breaking}

\title[A new eclipsing polar]
{Discovery, observations and modelling of a new eclipsing polar: \sourceNameLong} 

\author[Breytenbach et al.]{
  H. Breytenbach$^{1,2}$\thanks{E-mail: hannes@saao.ac.za}\,\orcid{0000-0001-5391-2386}, 
  D.A.H. Buckley$^{1,3}$\,\orcid{0000-0002-7004-9956},  %
  P. Hakala$^{4}$,
  J.R. Thorstensen$^{5}$\,\orcid{0000-0002-4964-4144},   %
  \newauthor 
  A. Y. Kniazev$^{1,3,6}$,
  M. Motsoaledi$^{1,2}$,
  P.A. Woudt$^{2}$\,\orcid{0000-0002-7004-9956},
  S.B. Potter$^{1}$\,\orcid{0000-0002-5956-2249},
  \newauthor 
  V. Lipunov$^{7,8}$, 
  E. Gorbovskoy$^{8}$,
  P. Balanutsa$^{8}$
  and N. Tyurina$^{8}$
  \\
  $^1$ South African Astronomical Observatory, PO Box 9, Observatory 7935, Cape Town, South Africa \\
  $^2$ Department of Astronomy, University of Cape Town, Private Bag, Rondebosch 7700, Cape Town, South Africa\\
  $^3$ Southern African Large Telescope, PO Box 9, Observatory 7935, Cape Town, South Africa \\
  $^4$ Finnish Centre for Astronomy with ESO (FINCA), University of Turku, V\"ais\"alantie 20, 21500 Piikki\"o, Finland	\\
  $^5$ Department of Physics and Astronomy, Dartmouth College, Hanover, NH 03755, USA \\
  $^6$ Special Astrophysical Observatory of RAS, Nizhnij Arkhyz, Karachai-Circassia 369167, Russia \\
  $^7$ M.V. Lomonosov Moscow State University, Physics Department, Leninskie gory, GSP-1, Moscow, 119991, Russia \\
  $^8$ M.V. Lomonosov Moscow State University, Sternberg Astronomical Institute, Universitetsky pr., 13, Moscow, 119234, Russia\\ 
}	

\date{Accepted XXX. Received YYY; in original form ZZZ}

\pubyear{\the\year}

\begin{document}
  \label{firstpage}
  \pagerange{\pageref{firstpage}--\pageref{lastpage}}
  \maketitle
  
  \begin{abstract}
    We report the discovery of a new eclipsing polar, \sourceNameLong, detected as an optical transient by MASTER auto-detection software at the recently commissioned MASTER-SAAO telescope. Time resolved (10--20 s) photometry with the SAAO 1.9-m, and 1.0-m telescopes, utilizing the SHOC EM-CCD cameras, revealed that the source eclipses, with a period of 2.08 hours (7482.9$\pm$3.5$\,$s). The eclipse light curve has a peculiar morphology, comprising an initial dip, where the source brightness drops to $\sim$50\% of the pre-eclipse level before gradually increasing again in brightness. A second rapid ingress follows, where the brightness drops by $\sim$60--80\%, followed by a more gradual decrease to zero flux. We interpret the eclipse profile as the result of an initial obscuration of the accretion hot-spot on the magnetic white dwarf by the accretion stream, followed by an eclipse of both the hot-spot and the partially illuminated stream by the red dwarf donor star. This is similar to what has been observed in other eclipsing polars such as HU~Aqr, but here the stream absorption is more pronounced. The object was subsequently observed with \SALT using the \gls*{RSS}. This revealed a spectrum with all of the Balmer lines in emission, a strong \HeII 4686~\AA{} line with a peak flux greater than that of H$\beta$, as well as weaker \HeI lines. The spectral features, along with the structure of the light curve, suggest \sourceNameLong is a new magnetic cataclysmic variable, most likely of the synchronised Polar subclass.%
    %
    
  \end{abstract}
  
  
  \begin{keywords}
  cataclysmic variables, accretion, magnetic fields, binaries: close, binaries: eclipsing, stars: individual: \sourceNameLong
  \end{keywords}
  
  

  \section{Introduction}
  \MCVs are close binary stars consisting of a magnetic \WD accreting from a low mass M or K dwarf companion \citep{Warner2003, Hellier2001}. The sub-class of \mCVs known as polars \citep{Cropper1990}, or sometimes AM Herculis stars, have white dwarfs whose rotation is synchronised (or nearly synchronised), to the orbital period. \MCVs are significant accretion-driven X-ray sources, with luminosities in the range of $L_X \sim 10^{31}-10^{32}$~erg~s$^{-1}$, a fact which has lead to many of them being discovered by X-ray surveys (see for example \citet{Rojas2017A&A...602A.124R},  \citet{Masetti2013A&A...556A.120M} and associated articles). 
  In polars, the strength of the white dwarf's magnetic field, coupled with the short orbital period -- and therefore close proximity of the two stars -- leads to accreted material from the inner Lagrangian point, L1, being injected near the magnetosphere of the primary.  At some distance from the white dwarf, the magnetic force dominates over the ram pressure of the accreting material, preventing the formation of an accretion disk.  Material then accretes along magnetic field lines, which funnel the plasma into an accretion column near the white dwarf's magnetic pole. As material impacts the surface of the \WD at supersonic velocities, a deceleration shock is produced in the flow at ${\sim}10^4$~m above the \WD surface. In passing through the shock, the bulk kinetic motion of the gas becomes thermalised, producing a hot (${\sim}10^8$~K) settling flow in the post-shock region. Some polars accrete onto both poles \citep[\eg][]{1995MNRAS.273..742B, 1995A&A...293..764S, 1999ApJ...525..407S, O'Donoghue2006}, while others are single pole accretors \citep[\eg][]{1993A&A...271L..25S, 2002MNRAS.331..488S, 2003MNRAS.341..863B}, and some switch between single and multiple pole accretion as the mass transfer rate or the magnetic field geometry changes \cite[\eg][]{1996MNRAS.280.1121R, 1998MNRAS.295..511M, 2016arXiv160908511D}.

  The X-ray spectra of \mCVs are typically characterised by two components, namely a hard ($kT > 10$~keV) bremsstrahlung component and a soft ($kT \sim 20-50$~eV) pseudo-blackbody component.  The former arises from bremsstrahlung emission in the post-shock accretion flow, just above the white dwarf surface. The latter is typically attributed to the result of reprocessing and/or direct bombardment due to a fragmented stream from which diamagnetic accretion ``blobs'' bury themselves in the photosphere of the white dwarf and re-radiate their kinetic energy \citep{Beuermann04}. 
  A resulting hot-spot is therefore produced near the magnetic poles which can often be seen in high time resolution eclipse light curves of polars (\eg \cite{O'Donoghue2006}). These hot-spots are typically very small, ${\sim}10^{-4}$ of the surface area of the white dwarf, but can produce a significant fraction of the total luminosity of the system. See \citet{Mukai2017PASP..129f2001M} for a review of the X-ray properties of accreting \glspl*{WD}.
  
  A defining characteristic of polars -- as their name suggests -- is the high level of optical polarization arising from cyclotron emission in the post-shock accretion flow, which can reach levels of ${\sim}50\%$ for circular polarization \citep{Hakala1994MNRAS.271L..41H}. Magnetic field strengths in polars, which range from 7--230~MG \citep{Ferrario2015}, have, for the most part, been determined by fitting models to spectro-polarimetric observations. In some cases, particularly when the accretion rate is low, magnetic field estimates can be made from measuring the Zeeman splitting of photospheric spectral lines \citep{Beuermann2007}. Another observational feature of polars is that their spectra typically show strong Balmer, \HeI and \HeII emission lines, with the latter generally much stronger ($F_{H\beta} < F_{\HeII}$) in polars and \mCVs compared to non-magnetic cataclysmic variables \citep{Szkody1998ASPC..137...18S}.  
  
\begin{figure}
    \includegraphics[width=0.48\textwidth]{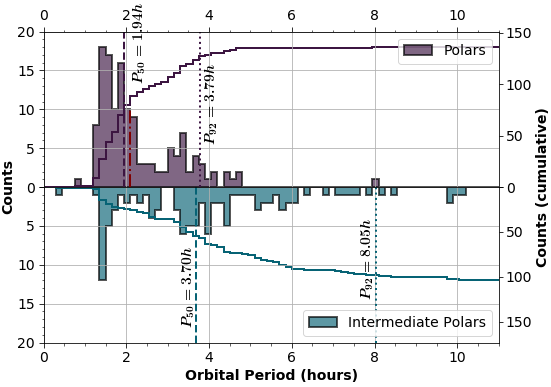}
    \caption{
      The orbital period distribution of the known \mCVs at the time of publication using data from \citet{Ritter2003}. The figure shows back-to-back histograms of the two classes of \mCVs over the orbital period range of 0--11~hours. Also shown for each class are the cumulative histograms (solid, stepped lines), the median period $P_{50}$ (dashed vertical lines), and the \nth{92} percentile period $P_{92}$ (dotted vertical lines).  The surplus of polars at short periods is clearly seen -- approximately 90\% of polars have periods shorter than the median period for the \IPs.  For reference, the orbital period of \sourceName (see \autoref{sec:period}) is also indicated as the red dash-dotted vertical line at $P=2.078$~h in the upper panel.
    }
    \label{fig:histogram}
  \end{figure}

  \Glspl*{mCV} constitute a small fraction of the total number of known CVs (${\sim}15-25$\%) -- about half of these being polars, and the other half the lower field asynchronous \IPs (for a review of \IPs, see \cite{Patterson1994PASP..106..209P}). 
  Of the 141 polars in the latest 
  edition of the Ritter~\&~Kolb catalogue of cataclysmic variables and related objects (v7.24, December 2015: \citet{Ritter2003}), 92\% have periods $\leq$3.8~h, with the shortest period system at 48.6~minutes.  The median period is 1.93~h, which means that slightly more than half of the known polars have periods below the ${\sim} 2-3$ hour period gap. \autoref{fig:histogram} compares the orbital period distribution of polars and \IPs. The surplus of the number of polars at shorter periods, as compared to that of the \IPs, can readily be seen. Whether or not the reason for this is due to an evolutionary transition or selection effects, remains a matter of debate \citep{Mason2004RMxAC..20..180M, 2012ApJ...746..165H}.   
  
  At the time of writing there are 33 polars that have been positively identified as eclipsing systems, which makes the discovery of a new example an important addition. Eclipsing systems allow indirect mapping of the gas flow and accretion on to the white dwarf, since the positions and dimensions of the accretion regions (the ``hot-spots''), the accretion stream and the interaction/threading region in the magnetosphere, can be determined by virtue of the secondary's occulting limb \citep{1995A&A...296..164H, Harrop-Allin1999, O'Donoghue2006}. In a wider context, eclipsing polars are ideal cosmic laboratories for studying magnetic accretion, which occurs in many other astrophysical objects, including protostellar objects (\eg T~Tau stars), pulsars and active galactic nuclei. 
  
  Here we report on the discovery of a new eclipsing cataclysmic variable from the MASTER-SAAO optical transient survey, \sourceNameLong (hereafter \sourceName). Follow-up time resolved photometry shows the system exhibits a deep, sharp eclipse, and a pre-eclipse ``dip'', likely due to obscuration of the accretion spot(s) by the accretion stream. Spectroscopy from \SALT reveals a spectrum characteristic of a polar, with the high-excitation emission line of \HeII at 4686~\AA{}, exceeding in flux that of H$\beta$ at 4861~\AA{}, a standard hallmark of an \mCV \citep{Szkody1998ASPC..137...18S}.

  \section{The MASTER--SAAO facility and the discovery of \sourceName}
  
  \subsection{MASTER--SAAO}
  \begin{table*}
    \caption{
      MASTER and CTRS survey parameter comparisons.
      }
    \label{tbl:surveys}
    
    \smallskip
    \begin{center}
      {\small
	\begin{tabular}{lcccc}  
	  \hline
	  \noalign{\smallskip}
	  Parameter 			& CRTS 				& CRTS 		& MASTER 		& MASTER 			\\
							& all nodes 		& SSS node 	& all nodes		& SAAO node 			\\
	  \noalign{\smallskip}
	  \hline
	  \noalign{\smallskip}
	  Collecting area (m$^2$) 			& 2.35 				& 0.20 				& 1.50 					& 0.25 							\\
	  Field of View (deg$^2$) 			& 13.3 				& 4.0 				& 64.0 					& 8.0 							\\
	  Maximum  area/night (deg$^2$) 	& $\sim$2500: 		& unknown 			& $\sim$4000 			& $\sim$500--1000 				\\
	  Cadence (days) 					& 7--10 			& 7--10 			& 10--15 				& 10--20 						\\
	  Exposures (sec) 					& 4 $\times$ 30 	& 4 $\times$ 30 	& 3 $\times$ (60--180) 	& 3 $\times$ (60--180)$^{*}$ 	\\
	  Limiting magnitude 				& 19--20; 22$^\#$ 	& 19--20 			& 18--20 				& 19--20 						\\
	  Alert threshold (mags) 			& 0.65 				& 0.65 				& 2.2					& 2.2 							\\
	  \noalign{\smallskip}
	  \hline
	  \noalign{\smallskip}
	  * depending on Moon phase \\
	  \# limit for CSS node in USA \\
	  \noalign{\smallskip}
	  
	  \hline\
	\end{tabular}
      }
    \end{center}
    
  \end{table*}
 
  \begin{figure*}
    \centering
    \includegraphics[width=1\textwidth]{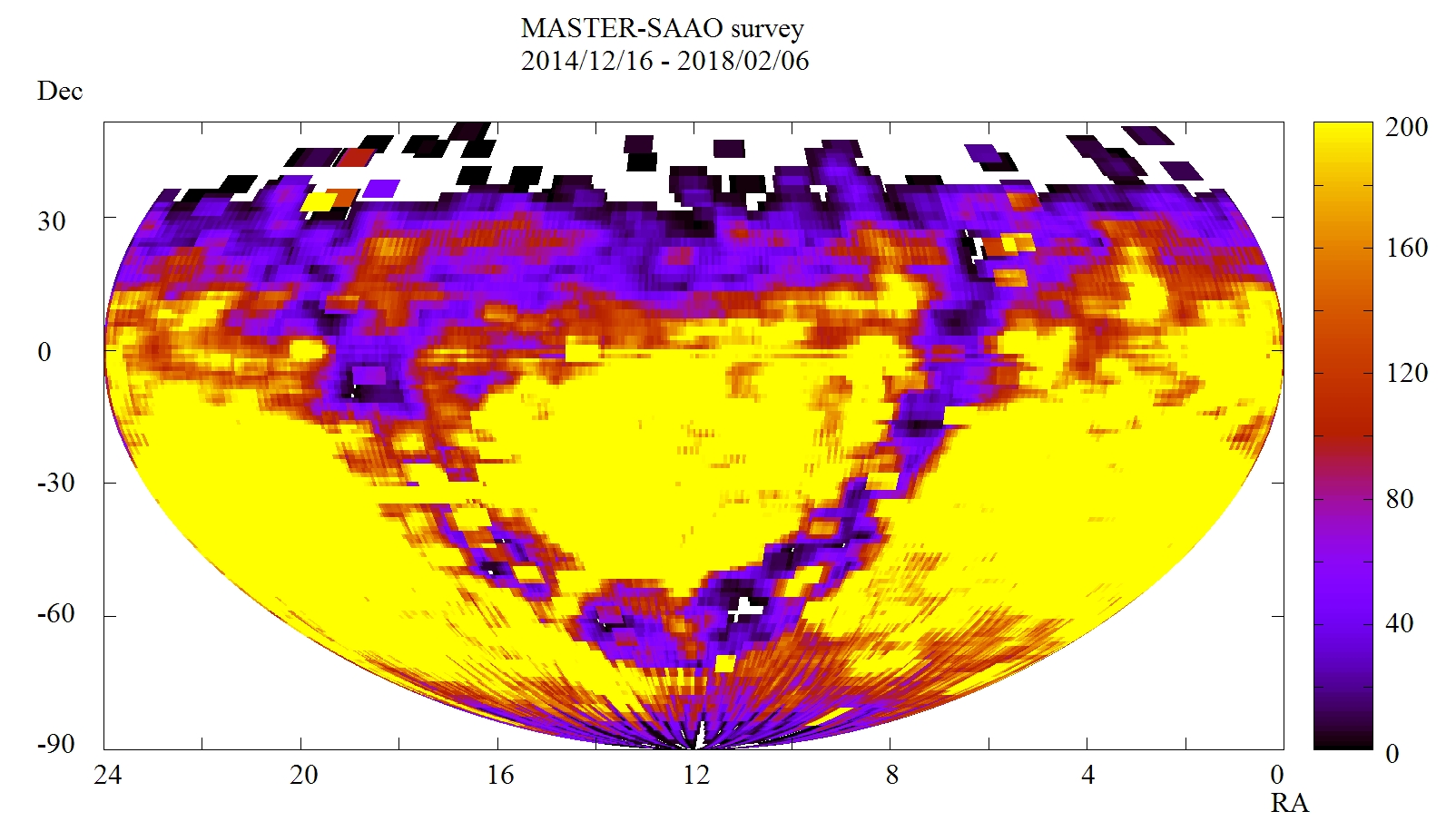} 
    \caption{
      The MASTER--SAAO survey area, colour coded to indicate the number of visits per field for the period 2014 December to 2018 February.
      }
    \label{fig:coverage}
  \end{figure*}
  
%
  The \MASTER\footnote{\url{http://observ.pereplet.ru/}} \citep{Lipunov2010AdAst2010E..30L, Kornilov2012ExA....33..173K} is a network of optical transient detection systems, the initial five situated in Russia, with three more recent nodes established elsewhere: MASTER-SAAO at the \SAAO, MASTER-IAC in Spain at Instituto de Astrof\'isica de Canarias (IAC, Tenerife) and MASTER-OAFA at the Observatorio Astronomico Felix Aguilar of San Juan National University in Argentina \citep{Lipunov2016RMxAC..48...42L}.

  The MASTER survey  detects a range of transient types, including Galactic sources (CVs, flare stars, novae, deeply eclipsing binaries), AGN (blazars), supernovae and Solar System objects. All the MASTER nodes can also undertake follow-up on transient alerts from other facilities, for example detecting the optical emission from GRBs or optical counterpart searches of gravitational wave or neutrino events. Each telescope can observe with one of four filters ($\it{B, V, R}$ or $\it{I}$) or with polarizing filters. As well as the twin 0.4-m telescopes, each mount also has 2 \gls*{VWF} cameras, covering $\sim$1000 deg$^2$, but with a shallow limit of unfiltered ${\sim} 12\,$mag in a 1 s exposure, and $15\,$mag for a 10~min exposure.
    
  The MASTER-SAAO node was installed and commissioned over a $\sim$3~week period, becoming operational in late December 2014.  A clam shell dome allows rapid slewing over the entire sky, with declination limit of $-90\deg$ to $+40\deg$. There is no limit on the Galactic latitude, although the Galactic plane is a lower priority region for MASTER--SAAO, so the cadence is considerably less (by 10--20~times) than for higher Galactic latitudes. In \autoref{fig:coverage} we show the survey area of MASTER-SAAO, which also indicates the number of visits for each position in the survey region.
  
  In \autoref{tbl:surveys} we compare some of the MASTER survey parameters with those of the \gls*{CRTS}\footnote{\url{http://crts.caltech.edu/}} \citep{Drake2009}, which until recently has been the only other transient detection system for which a study of CVs has been reported in some detail by \citet{Drake2014}. We attempt to make a comparison between the entire MASTER and \gls*{CRTS} networks, as well as with the two systems specific to the southern hemisphere, namely the MASTER-SAAO node in South Africa and the Siding Spring Survey (SSS) of \gls*{CRTS} in Australia (which ceased operations in 2013). While we have attempted to present the best estimates of the survey parameters, there are some inevitable uncertainties due to lack of published information and the fact that some of the parameters are dependent upon season, Moon phase and observing conditions.
  
  
  \begin{figure}
    \centering
    \includegraphics[width=0.48\textwidth]{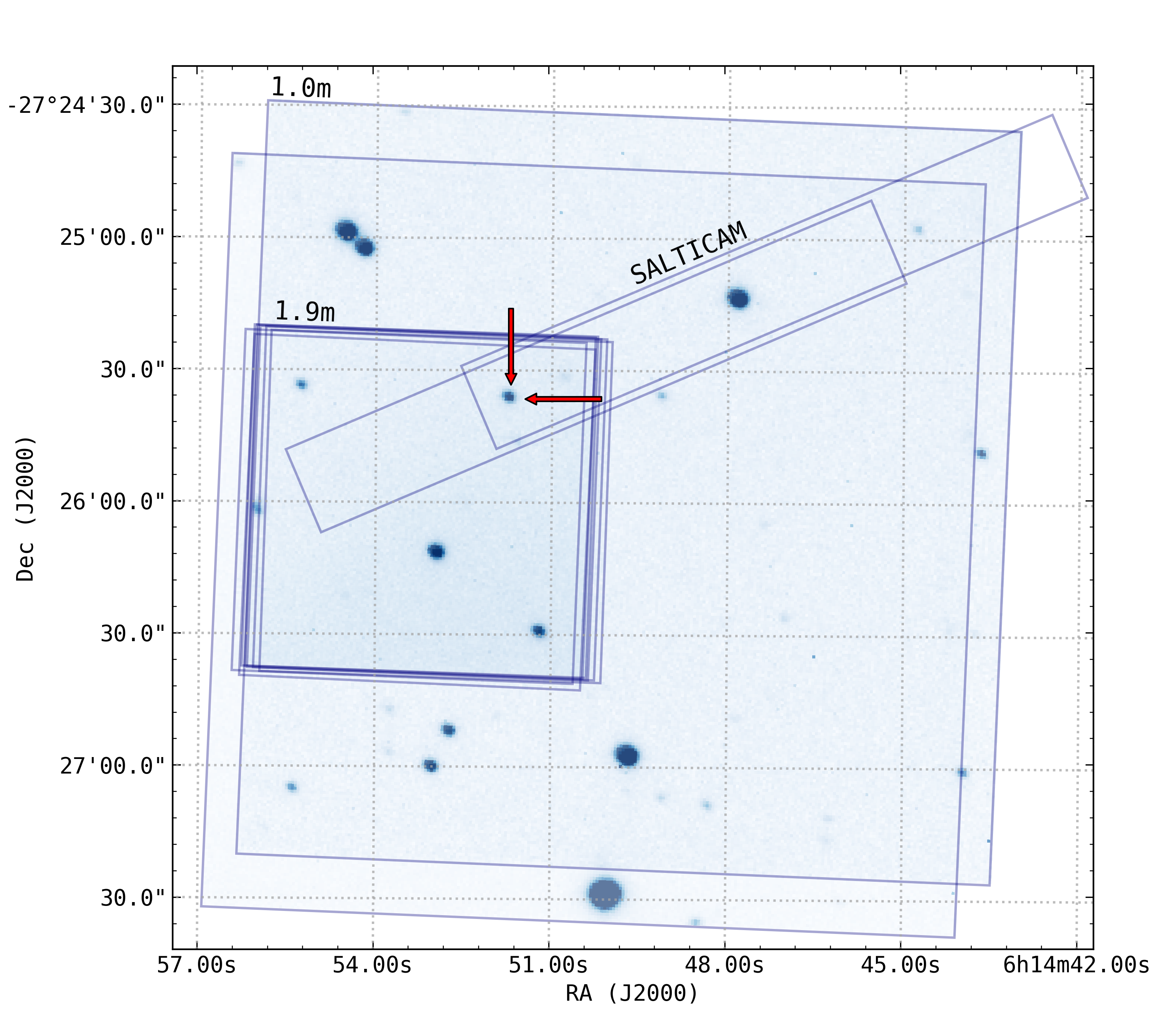}
    \caption{
      The on-sky instrument field-of-view in equatorial J2000 coordinates for the photometric data collected on \sourceName from the SAAO.  See \autoref{tbl:observations} for details.  The slight offset between images is due to manual telescope pointing. 
    }
    \label{fig:sky}
  \end{figure}

  \subsection{The discovery of \sourceName}
  
  The MASTER-SAAO auto-detection system \citep{Lipunov2010AdAst2010E..30L} discovered the transient source \sourceName on 2015 February 19 \citep{Shumkov2015ATel.7127....1S} 
  at 23h~17m~U.T at the position:
  \begin{ceqn}
    \begin{align*}
      \alpha_{2000} &= \hms{\ 06}{14}{51.70}\\
      \delta_{2000} &= \dms{-27}{25}{35.5}\\
    \end{align*}
  \end{ceqn}
  
  \begin{figure*}
    \includegraphics[width=0.9\textwidth]{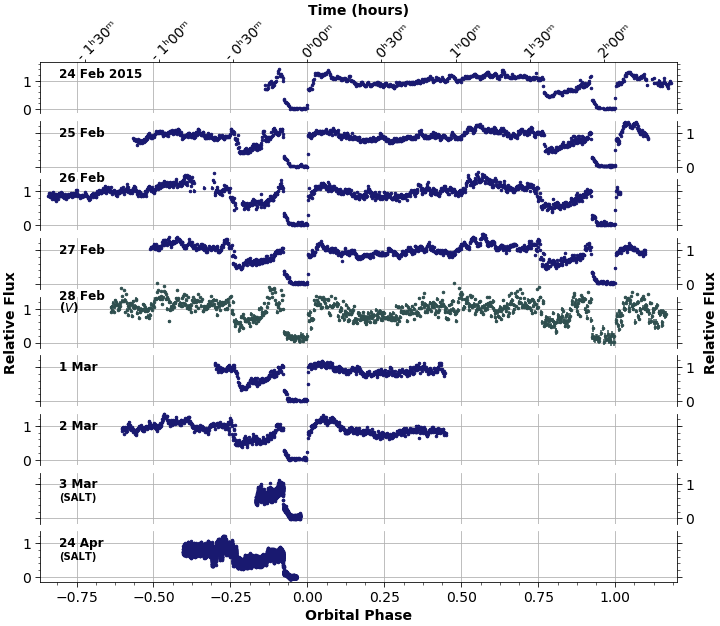}
    \caption{
      Light curves of \sourceName obtained in period 24 Feb -- 24 Apr 2015 using the SHOC high speed camera on the SAAO 1.9-m and 1.0-m telescopes (upper 7 light curves), as well as SALTICAM on SALT (lower 2 light curves). The eclipse and pre-eclipse dip, due to the obscuring accretion stream, are clearly seen.  All light curves are unfiltered (or clear filtered for SALT), save that of 28 Feb, which is $V$-band filtered.  Each light curve is displayed on the same scale in units of relative flux.
      }
    \label{fig:lightcurves2015}
  \end{figure*}

  The unfiltered magnitude was ${\sim} 18.3$ from the three consecutive images and the limiting magnitude of the reference image, taken on 2015 February 6.92896, was 19.4. There is one previous image of \sourceName in the MASTER-SAAO database, taken on 2015 January 10.996, where it was just detected at the 18.8 magnitude limit of the exposure.\footnote{The MASTER discovery images for \sourceName are available at \url{http://master.sai.msu.ru/static/OT/MASTERJ061451.70-272535.5.jpg}} 
  Further interrogation of this region with {\sc Aladin} has revealed a faint blue object, just above the limit, in the DSS UKSTU-blue survey image while there is also a weak {\it GALEX} source coincident with the \sourceName position. This source is catalogued as GALEX J061451.7-272534.
  
 \begin{table*}
    \centering
    \caption{Photometry observation log for \sourceName.} 
    \begin{tabular}{*{8}{c}}
      \hline \rule{0pt}{10pt}
      Date	& Time Start & Duration  & Exposure  & Filter & Telescope & Accretion & Figure \\
      		& (UTC) 	 & 			 & Time (s)  & 		  &			  & State			   \\
      \hline
      
      2015-02-24 	& 18:15:36 	& \hms{3}{26}{20} 	& 15 	& --		& SAAO 1.9m	& high 			& \ref{fig:lightcurves2015}	\\
      2015-02-25 	& 18:19:05 	& \hms{3}{28}{38}	& 10 	& --		& SAAO 1.9m	& high 			& \ref{fig:lightcurves2015}	\\
      2015-02-26 	& 18:41:14 	& \hms{3}{51}{59} 	& 10 	& --		& SAAO 1.9m	& high 			& \ref{fig:lightcurves2015}	\\
      2015-02-27 	& 18:14:32 	& \hms{3}{20}{48}	& 10 	& --		& SAAO 1.9m	& high 			& \ref{fig:lightcurves2015}	\\
      2015-02-28 	& 18:55:15 	& \hms{3}{45}{06}	& 15 	& V 	 	& SAAO 1.9m	& high 			& \ref{fig:lightcurves2015} \\
      2015-02-28 	& 18:40:16 	& \hms{4}{00}{02}	& 45 	& R 	 	& SAAO 1.0m	& high 			&	\\      
      2015-03-01 	& 18:29:20 	& \hms{1}{33}{23} 	& 10 	& --		& SAAO 1.9m	& high 			& \ref{fig:lightcurves2015}	\\
      2015-03-02 	& 19:16:22 	& \hms{1}{33}{23} 	& 10 	& --		& SAAO 1.9m	& high 			& \ref{fig:lightcurves2015}	\\
      2015-03-03 	& 20:37:31 	& \hms{0}{18}{58} 	& 0.5 	& clear 	& SALT		& high	  	 	& \ref{fig:lightcurves2015} \\
      2015-04-24 	& 17:16:26 	& \hms{0}{45}{41}	& 0.5 	& clear 	& SALT		& high 			& \ref{fig:lightcurves2015}	\\
      2017-02-08	& 19:16:22  & \hms{2}{30}{01}	& 45 	& --		& SAAO 1.0m	& low			& \ref{fig:folded}			\\
      2017-02-09	& 18:45:48 	& \hms{2}{30}{01}  	& 45 	& --		& SAAO 1.0m	& low			& \ref{fig:folded}			\\      
      2017-03-24 	& 02:36:58	& \hms{0}{24}{40}	& 12	& GG420		& MDM 1.3m	& low				\\
      2017-10-20	& 11:39:47	& \hms{0}{18}{26}	& 20 	& GG420		& MDM 1.3m	& low				\\
      \hline
					&		& \hms{29}{17}{25}										\\
      \hline
    \end{tabular}
    \label{tbl:observations}
  \end{table*}

  \section{High speed optical photometry}
  
  \subsection{Data acquisition and light curve extraction}
  \sourceName was observed at the \SAAO Sutherland station as part of a wider follow-up programme on new CV and transient discoveries resulting from the MASTER--SAAO survey (amongst others).  Time resolved (10--45 s resolution) photometry of \sourceName was obtained with both the \SAAO 1.9-m and 1.0-m telescopes during a two week campaign in 2015 February, with less extensive observations following up in 2017 February (see \autoref{tbl:observations} for a detailed observing log). Observations were undertaken with the \SHOC instruments, which utilise Andor iXon888 CCD cameras with 1024 $\times$ 1024 pixel CCDs \citep{Coppejans2013PASP..125..976C}. Light curves spanning 1.5--4.0 h were obtained on 7 different nights, mostly with a clear filter. The field of view was ${\sim}1.3' \times 1.3'$ for the 1.9-m observations and ${\sim} 2.8' \times 2.8'$ for the 1.0-m (see \autoref{fig:sky}). In addition, short time series were obtained in 2017 March and October using the 1.3m McGraw-Hill telescope at MDM observatory, the former with a SiTE 1024 $\times$~1024 CCD detector and the latter with an Andor Ikon camera similar to \SHOC. The MDM observations served to confirm the cycle count.
  
  In addition to the \SAAO and MDM data, photometry was done at a higher time resolution of 0.5~s, using SALTICAM in `slot mode' on \SALT during the nights of 3 March and 24 April 2015.  Reduction of the \SAAO \SHOC CCD data was performed using the \pySHOC\footnote{\url{https://github.com/astromancer/pySHOC}} library; an open-source, object-oriented 
  \python library 
  built off the standard \SHOC reduction pipeline. Calibration of the raw CCD frames, included subtraction of median bias and dark frames, as well as flat-field correction with median frames constructed from exposures of the uniformly illuminated evening twilight sky. Aperture photometry was used to extract the light curves of all stars in the calibrated science images. \pySHOC uses the {\sc photutils}\footnote{\url{https://photutils.readthedocs.io/en/stable}} library \citep{photutilsv3.0} as a backend for performing aperture photometry. The aperture positions were chosen by tracking the average centroid positions of the brightest stars in the frame, while the aperture sizes were chosen based on the dispersion (standard deviation / FWHM) of a Gaussian \gls*{PSF} fitted simultaneously to all stars using a least-squares optimization. In this way, the aperture size adjusts to any changes in seeing that occur during the run to ensure optimal signal-to-noise extraction. Sky annuli for background subtraction were similarly scaled.  Differential photometry was done on the background subtracted light curves by co-adding the signals of all non-variable stars in the field into a single time series which was used to compute the differential flux of \sourceName.  Finally, points in the light curves with large uncertainty (due to inferior atmospheric conditions or too-bright background) were removed from the final product light curve shown in \autoref{fig:lightcurves2015}.
  
  Extraction of the SALTICAM light curves required a number of additional steps.  In particular, the SALTICAM images are affected by vignetting from the slit mask, as well as by PSF degradation due to drift in the primary mirror alignment on SALT (active, closed-loop alignment of mirror segments was not operational at the time of the observation). Since our target star was positioned near the edge of the frame for at least one of the SALTICAM runs, we modelled the background vignetting pattern with a smooth, piecewise polynomial in order to remove its effect and produce a more accurate flux measurement. Additionally, the comparison star was (unintentionally) positioned on a region of the CCD containing a number of bad pixels which have a lower response than their neighbours.  If uncompensated for, the small frame-to-frame dither during the observation leads to spurious variation that is purely instrumental in origin being introduced by these pixels.  Since no calibration flats are available for SALTICAM, an artificial flat field image was constructed from the last ${\sim} 120$ frames of the observation (off-target sky images) by comparing the value of each pixel in the array with the median value of its 8 nearest neighbour pixels, and averaging this ratio across all available frames. The flat field image produced in this way effectively normalised the pixel response and improved the quality of the photometry.  To compensate for the PSF degradation, an optimization step was performed at each image in order to select the highest SNR aperture for the stars.  This step involves maximizing the totalled SNR for elliptical apertures across 3 parameters (height, width and rotation) given the data. The aperture fluxes obtained from the optimal apertures were then processed as above to obtain the final SALTICAM light curves.

  \section{Light curve characteristics}
  \label{sec:lc}

  In \autoref{fig:lightcurves2015} we show all the differential light curves we obtained during our 2015 campaign on \sourceName.  The most prominent feature in the light curve is a deep eclipse during which the source brightness decreases to the level of the background sky. In addition, there is a shallower eclipse-like feature, or dip, preceding the primary eclipse and lasting ${\sim}4$~min. This pre-eclipse dip comprises an initial sharp drop in brightness to ${\sim}50$\% of the out-of-eclipse level, followed by a slower recovery back to nominal brightness. The rapid ingress of the primary eclipse then occurs, followed by an approximately linear decline of intensity over ${\sim}3$~min into a total eclipse. A rapid egress occurs some ${\sim}6.5$~min later. The features immediately following the primary egress remain highly variable from night to night. Sometimes a stepped feature can be seen $20--30\,$s after the start of primary egress. This is most clearly visible in the light curves on 2015 February~24 (both eclipses: E0, E1) and February~25 (second eclipse: E13) (see Figures \ref{fig:lightcurves2015} and \ref{fig:eclipses}).  In general, the post-egress light curve tends to gradually increase in brightness towards a level comparable to that of pre-eclipse. The exceptions being that of 2015 February~25 (E12) and March~1 (E70) where the gradual brightening phase is much less pronounced, and the egress step brings the source almost directly back to the pre-eclipse level.
  
  We identify the components of the light curve as follows: The pre-eclipse dip is most likely due to an accretion stream obscuring a bright accretion spot(s) on the white dwarf, while the deep eclipse evidently arises from the secondary star passing in front of this bright spot. The residual brightness remaining after the primary eclipse can be attributed to the illuminated portion of the accretion stream which still remains visible beyond the limb of the secondary star. Similarly, the post-egress recovery, is most likely due to the egress of the illuminated accretion stream. Superficially, the light curve of \sourceName appears very similar to that of HU~Aqr presented in \citep{Harrop-Allin1999}, the most distinct difference being the prominence of the pre-eclipse dip in \sourceName.
  
  There is some variability in the phase of the pre-eclipse dip: A nightly shift is typically around 0.03 in phase, corresponding to an angular variation of ${\sim}10\deg$ of the obscuring portion of the stream with respect to the bright spot.  Comparing the light curves in \autoref{fig:folded}: The 2017 data were acquired while the source was in a lower accretion state and therefore appeared significantly fainter for most of the orbit. There are a number of interesting differences between the low- and high-state light curves, 
  most notably the phasing of the pre-eclipse dip.  
  The differences between the low- and high-state light curves are discussed in \autoref{sec:discuss} below.    
  
    Stochastic flickering, with an amplitude as much as ${\sim}20$\% of the peak intensity, is seen in the light curves, with a characteristic timescale of ${\sim} 200\,$s. This is particularly evident at $0.5 \lesssim \phi  \lesssim 0.8$ (\eg on February 26 \& 27), prior to the dip. In contrast to the high-frequency flickering often observed in polars with timescales of seconds (\eg VV~Pup), the flickering in \sourceName occurs on a time-scale of minutes.  Our data lack the time resolution to search for signatures of other rapid variability such as \qpos that occur in some \mCVs during their bright phases (\eg V834~Cen \citep{2017A&A...600A..53M}).
  
 \subsection{Period analysis}
  \label{sec:period}
  The most precise phase fiducial in the light curve (detailed in the next section) appears to be the sharp egress from eclipse. Since the SALTICAM light curves do not cover the eclipse egress, they were excluded for the ephemeris calculation. The remaining observations define a unique ephemeris,
  \begin{ceqn}
  	\begin{equation}
      \label{eq:ephem}
      \mathrm{BJD\, (egress)} = 2457078.2747(1) + 0.08660831(2) \, E
    \end{equation}
  \end{ceqn}
  where $E$ is the integer cycle count%
  \footnote{The timings are referred to the solar system barycentre and the UTC system.  To sufficient accuracy, our timings can be converted to TDB by adding 68.18~seconds to the 2015 data and 69.18~seconds to the 2017 data.}. %
  \autoref{tbl:ephemeris} gives timings of the sharp egress; the predicted times are computed using the ephemeris above.  The cadence of our data is insufficient to resolve the egress, so we are not able to estimate meaningful uncertainties for the egress times from the data alone, save to say that we are confident we can interpolate the half-egress point to an accuracy significantly better than our sampling interval. The uncertainty in the measured egress timings is estimated from the residuals of the linear ephemeris.  We have 12 egress timings, spanning two years, with an rms scatter of $5.4$~sec, which is half of our typical cadence.
  
  \begin{table}
    \centering
    \caption{Ephemeris of eclipse egress times for \sourceName. Observed/predicted times given are full Julian days minus 2450000.  The second-to-last columns gives the O$-$C~error in units of seconds, while the last column gives that in units of $\sigma_E$ - the estimated standard deviation uncertainty of the ephemeris.}
    \label{tbl:ephemeris}
    \begin{tabular}{r l l r r} 
      \hline \rule{0pt}{10pt}
      $E$  	&  Observed  	&  Predicted   	& \multicolumn{2}{c}{O$-$C} 	 		\\
      		&				&				& s			& $\sigma_E$				\\
      \hline
	   0  	&  7078.2748  	&  7078.2747  	&     4 	& 0.5						\\
	   1  	&  7078.3614  	&  7078.3613  	&     5 	& 0.6 						\\
	  12  	&  7079.3140  	&  7079.3140  	&  $-$1 	& 0.1 						\\
	  13  	&  7079.4005  	&  7079.4006  	& $-$12 	& 1.4						\\
	  35  	&  7081.3060  	&  7081.3060  	&  $-$3 	& 0.4 						\\
	  36  	&  7081.3927  	&  7081.3926  	&     3 	& 0.4						\\
	  58  	&  7083.2981  	&  7083.2980  	&     3 	& 0.4						\\
	8256  	&  7793.3129  	&  7793.3128  	&     9 	& 1.0 						\\
	8257  	&  7793.3994  	&  7793.3994  	& 	0 		& 0.0 						\\
	8268  	&  7794.3521  	&  7794.3521  	&  $-$1		& 0.1 						\\
	8756  	&  7836.6173  	&  7836.6171  	&     1		& 0.1						\\
	11185 	&  8046.9884  	&  8046.9885  	&  $-$7		& 0.8						\\
      \hline
    \end{tabular}
 \end{table}
 
\begin{figure}
    \includegraphics[width=0.5\textwidth]{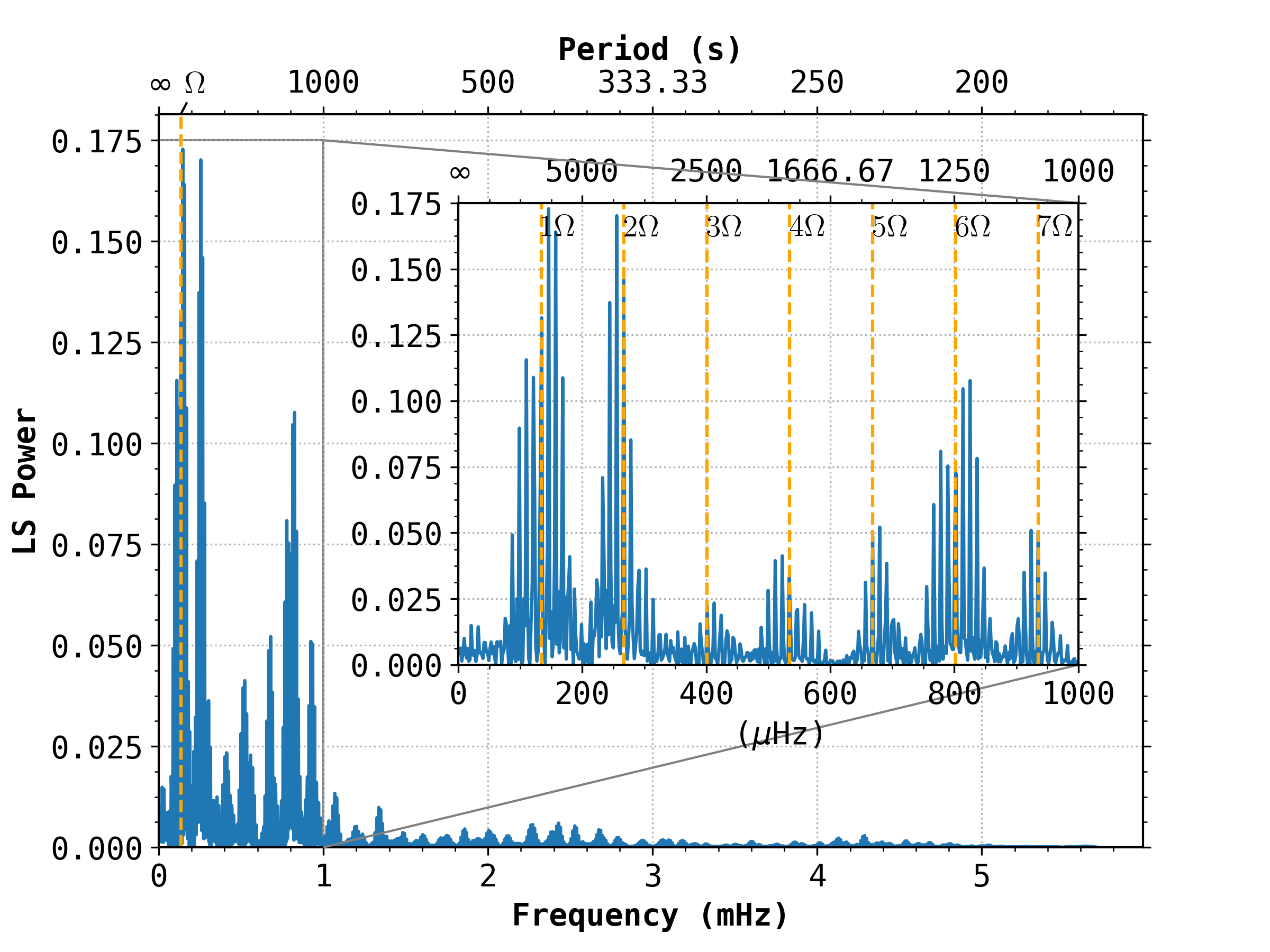}
    \caption{
      The Lomb-Scargle periodogram of the combined light curve data of \sourceName. The most prominent peak in the periodogram is due to the orbital variation and its harmonics (the orbital period is indicated by the dashed orange line). The inter-day alias structure due to the observing cadence is also clearly seen in the inset panel as the separation between adjacent peaks.
    } 
    \label{fig:periodogram} 
  \end{figure}

 Since the eclipse egress provides a well constrained marker for timing analysis, we do not need to rely on a frequency analysis to determine the orbital period. Nonetheless, we performed a frequency-spectral estimation using the Lomb-Scargle technique \citep{1976Ap&SS..39..447L, 1982ApJ...263..835S, VanderPlas2015ApJ...812...18V} in order to search for hidden periodicities, and possible quasi-periodic variability at higher frequencies. \autoref{fig:periodogram} shows the periodogram computed using the {\sc gatspy}\protect\footnote{\url{http://www.astroml.org/gatspy/}} \python package \citep{jake_vanderplas_2015_14833}. We see no indication of significant periodic variability not associated with the orbital variation and its harmonics.
 
\subsection{Measuring the eclipse profiles}
\label{sec:eclMeasure}
\begin{figure}
    \includegraphics[width=0.48\textwidth]{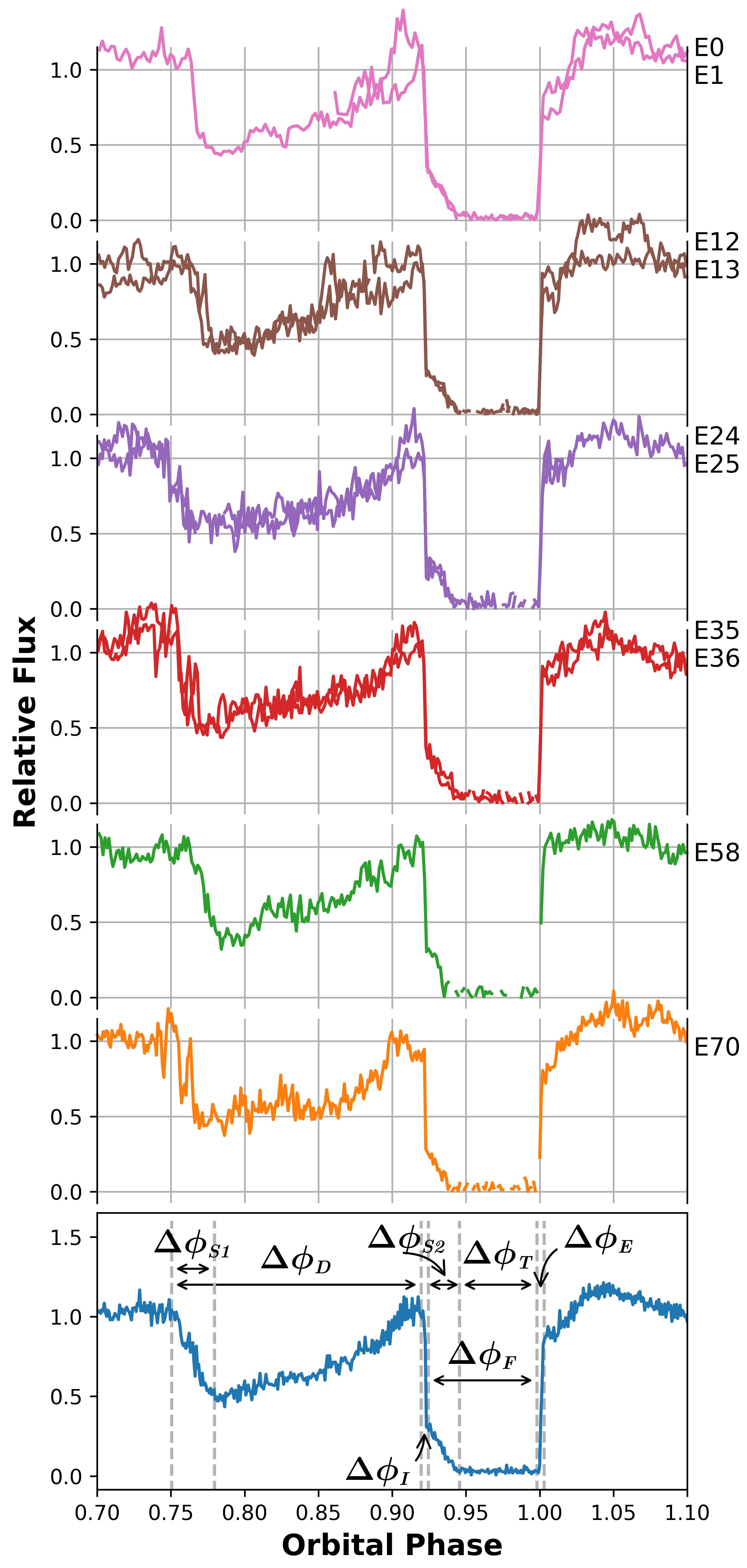}
    \caption{
      The 10 eclipses of \sourceName observed over 6~nights of observing during our 2015 campaign. The cycle counts from the ephemeris in \autoref{eq:ephem} are indicated next to the light curves on the right of the figure. The bottom panel shows the average light curve with the various phases of the eclipse marked and labelled (see also \autoref{tbl:eclipseParams}).
    }
    \label{fig:eclipses}
  \end{figure}

\autoref{fig:eclipses} shows the succession of the eclipse light curves for the 2015 data for a phase range of $0.7 < \phi < 1.1$.  The steep initial decline of the first dip, followed by a steady recovery in intensity, followed by the eclipse are clearly seen. We present the measured phase duration of various parts of the eclipse (labelled in the bottom panel of \autoref{fig:eclipses}) in \autoref{tbl:eclipseParams}. The values presented here were, in each case, obtained by measuring the durations of each component in the individual light curves (using only the 2015 data). The values in \autoref{tbl:eclipseParams} are the means of the individual measurements, while the uncertainties are their $1 \sigma$ standard deviations.\\
The width of the primary eclipse was measured as the full width at half depth (FWHD). The out-of-eclipse brightness used as a reference point to measure the depth of the eclipse is taken as the averaged brightness in phase bins of width 0.025 immediately preceding/following the primary eclipse. It was found that the resulting FWHD measurement is relatively insensitive to the choice of bin size for measuring the out-of-eclipse brightness.\\
To measure the duration of the ingress / egress, we refer to the point-to-point difference estimate for the light curve derivative. Due to the presence of flickering noise from the accretion process, it is not possible to distinguish the ingress/egress of the \WD as distinct from that of the accretion column / bright spot given our data.  We therefore refer to the combined observation of both these features as the primary ingress. The start of the primary ingress/egress is taken as the point preceding that at which there is a significant change in the derivative estimate. Here it proved sufficient to use the Generalized Extreme Studentized Deviate (GESD) Test \citep{Rosner83} to identify the outlying points marking the start of the ingress and egress. The end of the ingress/egress is taken as the point preceding that at which the point-to-point derivative again returns to nominal (inlying) values. Using this method to measure the ingress and egress duration from the data implies that the uncertainty on our measurements is at least as large as the time resolution (exposure time) of each light curve. We note furthermore that the size, shape, and position of the accretion column, as well as any brightness variability within it (which we know occurs on timescales shorter than the orbital period) will also affect these measurements. The time (phase) resolution of the light curves from the smaller telescopes (SAAO 1.0-m and 1.9-m) is insufficient to resolve the primary ingress/egress.  The two SALT light curves (2015 March 3 and April 24), however, do resolve the primary ingress, but where terminated mid-eclipse, and therefore do not resolve the egress. The egress duration presented in \autoref{tbl:eclipseParams} should therefore be interpreted as an upper limit only.\\
The ingress of the lower portion of the stream starts immediately after the primary ingress and ends at the start of totality.  The start of totality is taken as the time corresponding to the first data point which falls below $1 \sigma$ of the in-totality noise level. 
To measure the duration of the pre-eclipse dip, and the ingress duration of the upper portion of the stream, the point-to-point numerical derivative proved inefficient, due to the high variability in this part of the light curve. Instead, we use a smooth estimate of the derivative obtained by a total variation regularization procedure. A detailed implementation of this procedure is presented in \cite{Stickel2010}. This method is similar to the edge-preserving derivative estimator used by \cite{Spark+2015MNRAS.449..175S} to investigate the variability of the boundary layer in the eclipsing non-magnetic CV, OY~Car. We find that the start and end of the dip ingress can reliably be detected by selecting the phase regions in which the optimally smooth derivative remains negative and has magnitude greater than the local outlier (GESD) threshold as described above.\\
Finally, we note that, in terms of measuring the eclipse parameters, the influence of noise from short timescale variability may be mitigated by incorporating effect of flickering in the modelling procedure, as done in \eg \cite{McAllister+2017}, although doing so is beyond the scope of this work.

 \begingroup
 \renewcommand{\arraystretch}{1.5}
 \begin{table}
 	
    \centering
    \caption{Measured duration of various parts of the eclipse light curve of \sourceName.  The segments listed here are illustrated graphically in the bottom panel of \autoref{fig:eclipses}.}
    \label{tbl:eclipseParams}
    
	\begin{tabular}{ >{\raggedright}m{2cm} m{0.9cm}@{$\pm$\hspace{1pt}}m{0.8cm}  m{0.75cm}@{$\pm$\hspace{1pt}}m{0.4cm}  m{0.8cm}@{$\pm$\hspace{1pt}}m{0.4cm} }
      \hline \rule{0pt}{10pt}
      & \multicolumn{2}{c}{$\Delta\phi$} & \multicolumn{2}{c}{$\Delta \phi$ ($\deg$)}	& \multicolumn{2}{c}{$\Delta t$ (s)}      \\
      \hline  		
      Upper stream ingress ($\Delta\phi_{S1}$)		& \uNrTbl{0.029}{0.007}     & \uNrTbl{10.0}{2.5} 	& \uNrTbl{216}{50}      		\\
      Primary ingress ($\Delta\phi_{\mathrm{I}}$)  	& \uNrTbl{0.0007}{0.0001}  	& \uNrTbl{0.24}{0.05}	& \uNrTbl{5}{1}     	   	  	\\  
      Lower stream ingress ($\Delta\phi_{S2}$)  	& \uNrTbl{0.0233}{0.0012}  	& \uNrTbl{8.4}{0.4}		& \uNrTbl{174}{9}      			\\
      Totality  ($\Delta\phi_{T}$)        			& \uNrTbl{0.0548}{0.0014}   & \uNrTbl{19.7}{0.5}	& \uNrTbl{410}{10}    			\\
      Primary Egress ($\Delta\phi_{E}$)       		& \uNrTbl{0.0025}{0.0008}   & \uNrTbl{0.9}{0.3}		& \multicolumn{2}{l}{$19 \ ^*$}   		    	    \\
      Eclipse FWHD	($\Delta\phi_{F}$)				& \uNrTbl{0.0780}{0.0004}	& \uNrTbl{28.08}{0.14}	& \uNrTbl{583.8}{2.7}			\\
      \hline
      \multicolumn{2}{l}{* upper limits}
	\end{tabular}

\end{table}
\endgroup
  
    
  \begin{figure}
    \includegraphics[width=0.47\textwidth]{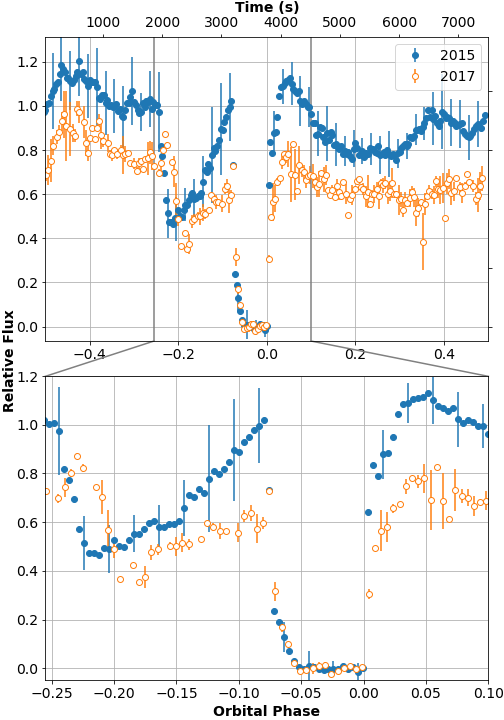}
    \caption{
      Average phase folded light curves of \sourceName comparing the high-state (filled blue circles) and low-state (open orange circles) runs. Uncertainties on the data points are indicated by vertical bars.  The source was significantly fainter in 2017.  The most prominent change in the eclipse profile between the two runs is the phase of the pre-eclipse dip, most likely due to changes in the location and/or radius of the threading region. The upper panel compares the light curve across the full orbital cycle, while the bottom panel shows a zoom-in of the eclipse.  The differences between the low-state and high-state light curves are discussed in the text of \autoref{sec:discuss}.
    }
    \label{fig:folded}
  \end{figure}
  

  \section{Spectroscopy}
\begin{figure*}
    \includegraphics[width=1\textwidth]{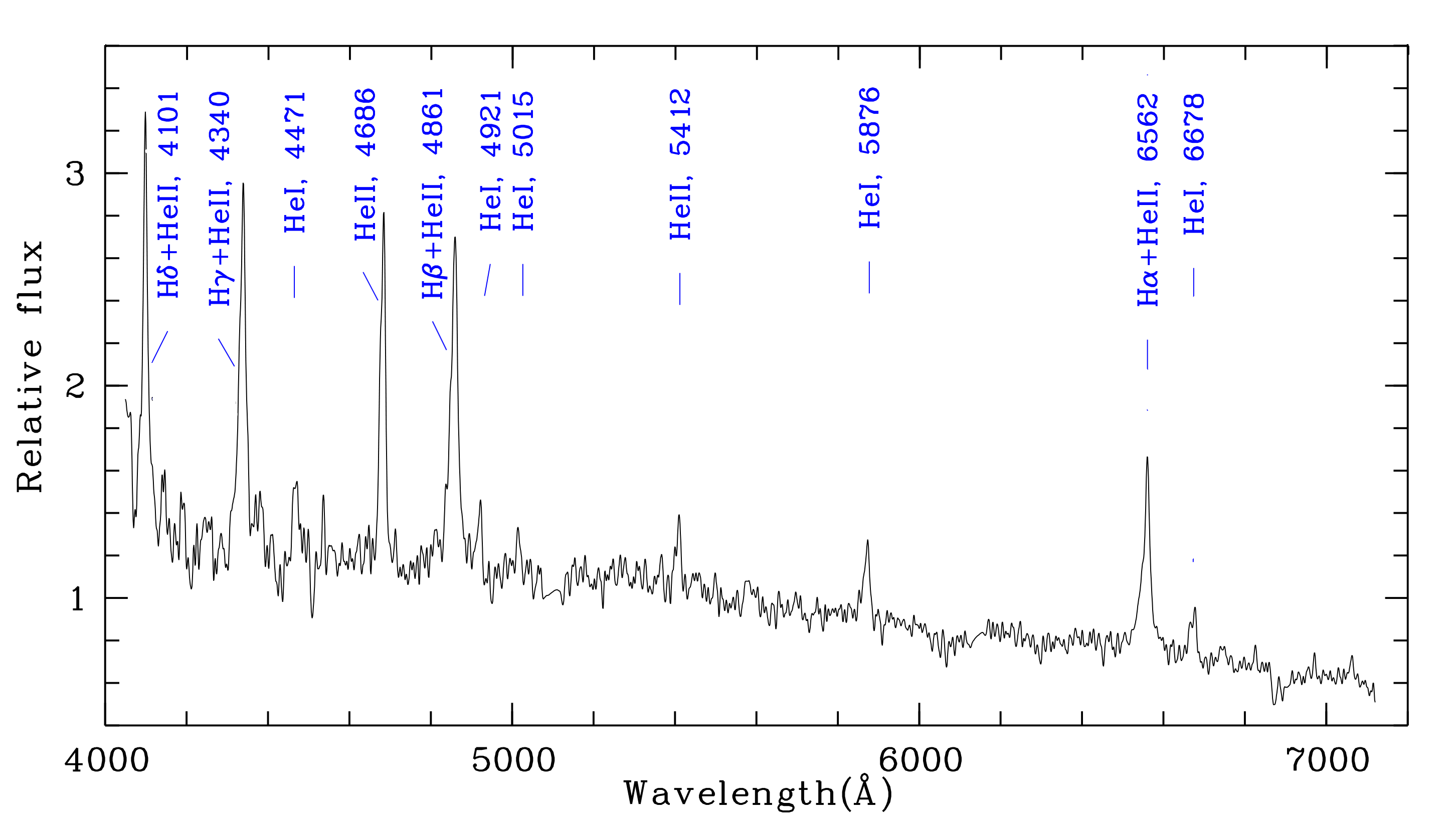}
    \caption {
      A SALT RSS spectrum of the eclipsing polar, \sourceName.
    }
    \label{fig:spectrum}
  \end{figure*}

  \sourceName was observed with the \SALT \citep{Buckley2006SPIE.6267E..0ZB} during our photometry campaign using the \gls*{RSS} \citep{Burgh2003SPIE.4841.1463B, Buckley2008SPIE.7014E..07B} on 2015 February 28 at 21:10~U.T., with the mid-time of the observation at HJD~=~2457082.3881. The PG900 grating and slit 1.25~arcsec width were used for observations. The spectrum covers spectral region 4060$-$7120~\AA{}, with a final reciprocal dispersion of $0.97$~\AA$\cdot$pixel$^{-1}$. The spectral resolution \gls*{FWHM} is $4.40\pm$0.15~\AA. The total exposure time was 800~s. An \textsc{Ar} arc lamp spectrum was taken immediately after science frame. Spectrophotometric standard stars were observed during twilight time for relative flux calibration. Absolute flux calibration is not feasible with SALT since the size of the unfilled entrance pupil changes during the observations as the telescope tracks. Primary reductions of the data was done in the standard way with the SALT science pipeline \citep{Crawford2010SPIE.7737E..25C}. We reduced the long-slit spectroscopic data using procedures described by \citet{Kniazev2008MNRAS.388.1667K}. The resulting normalized spectra are presented in \autoref{fig:spectrum}. 
  Equivalent widths (EWs), FWHMs and heliocentric radial velocities (RVs) of some lines in the spectra were measured applying the {\sc midas} programs (see \citet[][]{Kniazev2004ApJS..153..429K} for details) and are given in \autoref{tbl:spectrum}. The spectrum shows all of the Balmer lines in emission, a strong \HeII 4686~\AA{} line, at a peak flux greater than H$\beta$, and weaker \HeI lines, all superimposed on a continuum rising to the blue. The lines appear asymmetric and structured, with wings extending to shorter wavelengths. We note that the average radial velocity of the lines is -230~km~s$^{-1}$, consistent with the observation being at orbital phase $\phi = 0.49$, when the white dwarf is closest to us and the accretion stream from the secondary is expected to have a net blue-shifted velocity as it is approaching us.\\
  \\

 \begin{table*}
    \caption{Emission line parameters for \sourceName.}
    \begin{tabular}{llrrcr} 
      \hline \rule{0pt}{10pt}
      $\lambda_{0}$ (\AA) 	& Line			& E.W. (\AA)   		& FWHM (\AA) 		& R.V. (km s$^{-1}$)  	& Relative Flux \\ 
      \hline
      4101\ 			& H$\delta$ + \HeII  	& $18.67 \pm 0.66$ 	& $14.52 \pm 0.55$ 	&  -$242 \pm 17$ 		& $26.3 \pm 0.9$	\\
      4340\ 			& H$\gamma$ + \HeII  	& $28.68 \pm 0.83$ 	& $21.67 \pm 0.66$ 	&  -$209 \pm 19$ 		& $34.6 \pm 1.0$	\\
      4471\ 			& \HeI                	&  $6.06 \pm 0.46$ 	& $20.67 \pm 1.44$ 	&  -$200 \pm 40$    	&  $7.1 \pm 0.5$	\\
      4686\ 			& \HeII              	& $23.10 \pm 0.71$ 	& $15.81 \pm 0.50$ 	&  -$234 \pm 13$ 		& $27.0 \pm 0.8$	\\
      4861\				& H$\beta$ + \HeII   	& $28.83 \pm 0.84$ 	& $22.84 \pm 0.68$ 	&  -$239 \pm 18$ 		& $33.6 \pm 1.0$	\\
      4922\ 			& \HeI                  &  $2.77 \pm 0.36$ 	& $10.21 \pm 1.12$ 	&  -$214 \pm 29$    	&  $3.2 \pm 0.4$	\\
      5016\ 			& \HeI                  &  $1.73 \pm 0.27$ 	&  $9.17 \pm 0.96$ 	&  -$225 \pm 24$    	&  $2.0 \pm 0.3$	\\
      5412\ 			& \HeII              	&  $3.92 \pm 0.45$ 	& $13.95 \pm 1.37$ 	&  -$189 \pm 32$ 		&  $4.2 \pm 0.5$	\\
      5876\ 			& \HeI               	&  $6.84 \pm 0.53$ 	& $18.24 \pm 1.06$ 	&  -$263 \pm 23$ 		&  $6.2 \pm 0.5$ 	\\
      6563\ 			& H$\alpha$ + \HeII  	& $26.56 \pm 1.03$ 	& $24.00 \pm 0.85$ 	&  -$201 \pm 17$ 		& $20.4 \pm 0.8$	\\
      6678\ 			& \HeI               	&  $5.32 \pm 0.53$ 	& $19.24 \pm 1.44$ 	&  -$245 \pm 27$ 		&  $3.8 \pm 0.4$	\\
      \hline
    \end{tabular}
    \label{tbl:spectrum}
  \end{table*}
 
  The spectrum of \sourceName is typical of an \gls*{mCV}, particularly with the strong \HeII~4686~\AA{} line. Therefore, considering the spectral features together with the features in the light curve -- the rapid and deep eclipses as well as the pre-eclipse ``dip'' and the absence of other periodicities -- the evidence points to \sourceName being new eclipsing polar.

  \section{Modelling the eclipse and pre-eclipse dip}
  \label{sec:modelling}
  The light curve of \sourceName is characterised by a typical polar-like eclipse profile, where the eclipse is dominated by the rapid ingress and egress of the accreting pole on the WD surface.  This is preceded by a deep and very wide absorption feature or dip. This feature is much more prominent than usually seen in polars, as it is both much deeper and more extended in orbital phase. The eclipse of the accretion stream is characterised by the slow and linear transitions after the pole ingress and egress. The depth and substantial phase extent of the absorption dip implies somewhat unusual accretion geometry. In order to investigate the physical origin of the eclipse light curve, we have employed the eclipse mapping approach of \citet{Hakala2002MNRAS.334..990H}, which aims to provide three-dimensional mapping of the accretion stream and has previously been used to model the light curves of a number of other polars \citep{2004MNRAS.351.1423B}.
  
\citet{Hakala2002MNRAS.334..990H} constructed a gridless 3D model for the accretion stream, where the stream was modelled by a swarm of ``fireflies'' \ie emission points of equal brightness, that were free to move within the Roche lobe of the primary. As such the model has far too much freedom when compared to the information content of the data, and some regularisation is required.  The model was regularised by preferentially selecting firefly distributions that follow a smooth trajectory from the L1 point to the \WD surface. This was achieved by first solving the smoothest possible line through any swarm of fireflies using a \gls*{SOM} algorithm \citep{Kohonen1990}. The deviations of fireflies from this trajectory were then minimised, together with the $\chi^{2}$ for the fit (see the original paper of \citet{Hakala2002MNRAS.334..990H} for more details).
  
    
 
  \begin{figure*}
  	\includegraphics[width=1\textwidth]{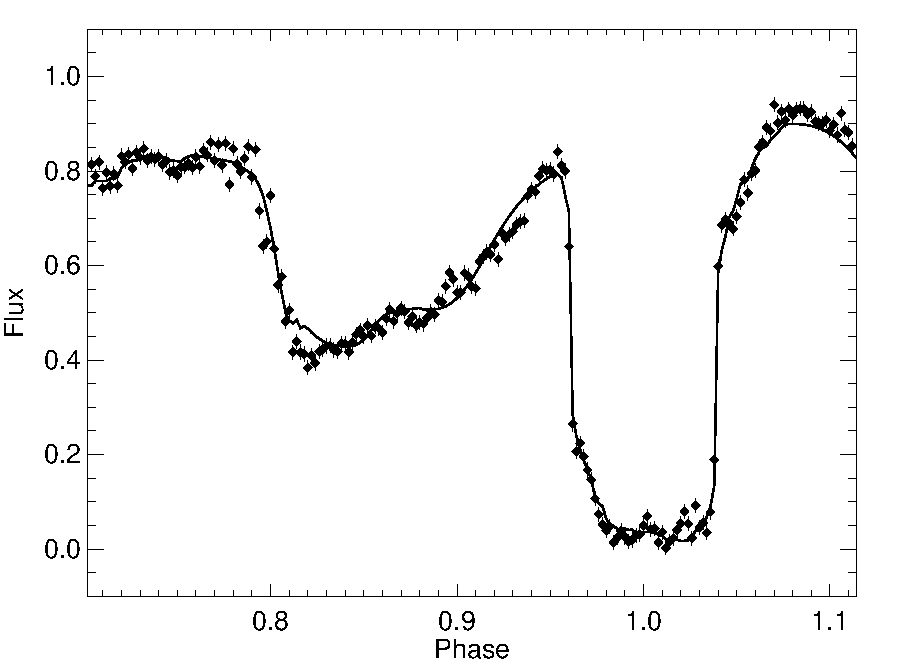}
    
    \includegraphics[width=1\textwidth]{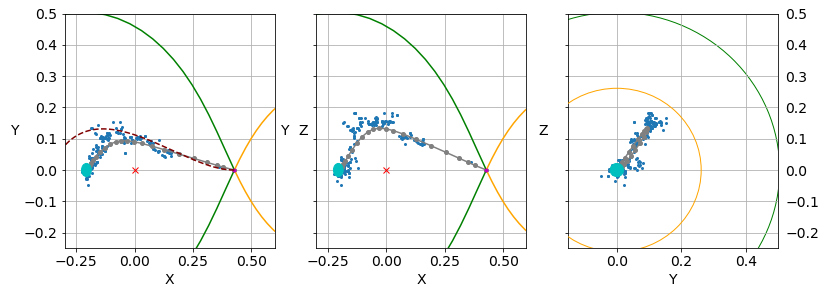}
    \caption[]{
    Upper panel: The accretion stream model fit to the average 2015 light curve of \sourceName. 
    Lower panel: A view into the accretion geometry. Each panel provides a two-dimensional projected view of the system in units of the binary separation $a$.  Shown in grey is the trajectory of the accretion stream derived from the ``firefly'' model.  The best fitting model shows the stream trajectory that moves above the orbital plane by up to $10 R_{\mathrm{WD}}$. The scattered blue dots around the final stream solution are the positions of the individual fireflies used to derive the fit. The position and radius of the \WD are indicated by the cyan dot. The centre of mass of the system (red cross), as well as the Roche lobes of both stars are also partially shown.  The leftmost panel (view from above the orbital plane) also shows a typical single particle ballistic trajectory for a hydrogen particle entering the system at the L1 point with thermal velocity typical of a late M--type dwarf at 3000~K.   The system parameters used to construct these plots are derived in \autoref{sec:params}.
    } 
    \label{fig:stream-model}
  \end{figure*} 
 
  \citet{Hakala2002MNRAS.334..990H} did not model any dips and therefore the original model did not include any absorption. Clearly absorption is required here, so the model was updated.  As the distribution of fireflies mimics the mass distribution of the stream, we implemented the first order model for the absorption by simply associating a 3D Gaussian optical thickness profile for each firefly and integrated the optical depth from the \WD along the line of sight. The self-absorption of the stream emission was not considered. We experimented with various Gaussian widths, but the results were not strongly dependent on the exact choice thereof.
  
  The orbital inclination,~$i$, is not known, but since the system is eclipsing, we may constrain it to be greater than $75\deg$. To further constrain the inclination, we used our code to compute the eclipse widths for a range of $(q, i)$ pairs. For each value of inclination we examined a grid of q values to reproduce the observed eclipse width for the pole. The results are tabulated in \autoref{tbl:inclination}. The $(q,i)$ values remain fixed during each eclipse profile modelling run.
  
  \begin{table}
    \centering
    \caption{Possible combinations of inclination ($i$) and mass ratio ($q$) pairs based on the observed eclipse width.}
    \begin{tabular}{l|*7{c}}
     \hline
     \hline
		$i$	& $76\deg$ 	& $78\deg$ 	& $80\deg$ 	& $82\deg$ 	& $84\deg$ 	& $86\deg$ 	& $88\deg$ \\
      \hline
		$q$	& 0.713 	& 0.543 	& 0.420 	& 0.332 	& 0.272 	& 0.233 	& 0.211 \\
      \hline
      \hline
    \end{tabular}
    \label{tbl:inclination}
  \end{table}
  
We find that best fits are obtained with intermediate inclination values around 80 degrees, even if our modelling cannot be used to constrain the inclination properly. This is because the $\chi^{2}$ values of different fits cannot be formally compared due to the regularisation of the model. The resulting light curve, together with the best fitting model is shown in \autoref{fig:stream-model}. The underlying 3D model of the accretion stream (distribution of fireflies) with its associated regularisation reference (20~nodes calculated using \gls*{SOM}) is shown in \autoref{fig:stream-model}. Our modelling suggests that the wide dip preceding the eclipse is likely to be produced by a relatively wide accretion stream/curtain in the magnetically controlled part of the stream near the \WD. The stream appears to be following a trajectory, where it is taken well above the orbital plane (up to ${\sim} 10R_{WD}$) soon after leaving the L1 point. The stream impacts the \WD at a ${\sim}45\deg$ angle with respect to the orbital axis, and a ${\sim}40\deg$ azimuthal angle, with respect to the line of centres. The horizontal and vertical extents of the stream are very similar. Most of the stream emission arises within $0.3\,a$ of the \WD, where $a$ is the binary orbital separation. 
    
  The results of the modelling are presented in \autoref{fig:stream-model} where we show in the upper panel, our model fit to the eclipse and the stream absorption feature. The model stream trajectory, as well as the firefly distribution from which it was derived, is show in the bottom panel of \autoref{fig:stream-model}.

  \section{Binary system parameters}
  \label{sec:params}
   
  Using the semi-empirical evolutionary models of \cite[supplementary Table 6]{Knigge2011ApJS..194...28K}, we estimate the donor mass and radius from the orbital period as $M_2 = 0.163 \Mo$ and $R_2 = 0.202 \Ro$. Since we are able to measure the orbital period of \sourceName very accurately (see \autoref{sec:period}), the uncertainties in $M_2$ and $R_2$ can be considered to be purely due to the evolutionary model (the propagated uncertainty due to the period measurement is less than 0.05\%).  We note that the Knigge $M_2-R_2$ relation assumes a non-magnetic primary with a mass of $0.75 \Mo$.  Any effects on the evolution of the system that are produced by magnetic activity (and/or rotation of the secondary) are therefore unaccounted for here.  Since the \textit{model} uncertainties are not made available in \cite{Knigge2011ApJS..194...28K}, it is difficult to assign uncertainties on the parameters $M_2$ and $R_2$. Based on the scatter of data points in the $M_2-R_2$~relation (Figure 4 in \cite{Knigge2011ApJS..194...28K}), we expect a ${\sim} 15\%$ uncertainty in these parameters.
  
\begin{figure}
    \includegraphics[width=0.48\textwidth]{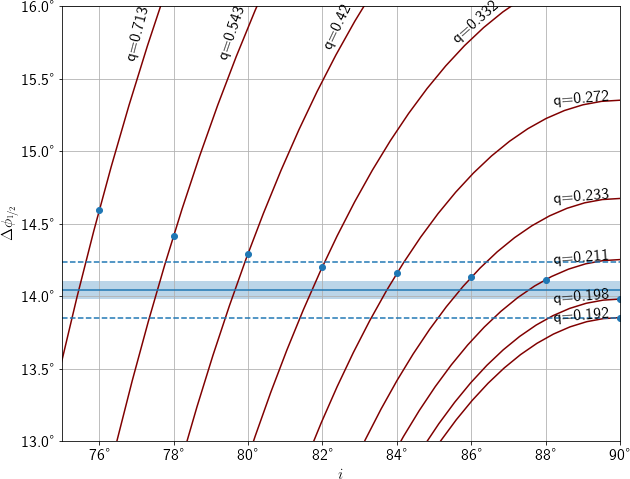}
    \caption{
    Constraints on the inclination of the system based on the eclipse phase width $\Delta\phi_{\nicefrac{1}{2}}$. Nine level contours of the $q(i, \Delta\phi_{\nicefrac{1}{2}})$ relation are shown as red curves crossing the axes diagonally.  For each contour, a circular marker indicates the inclination value estimated from the firefly model. The measured eclipse half width (solid horizontal blue line), as well as its 1$\sigma$ (shaded horizontal region) and 3$\sigma$ uncertainties (dashed horizontal blue lines) are also shown. 
   }
    \label{fig:iConstraint}
  \end{figure}

  Having measured the phase width ($\Delta\phi$) of the primary eclipse, we may place constraints on the inclination,~$i$, and mass ratio,~$q$, that are independent of the firefly model using the method of \cite{Chanan+1976ApJ...208..512C} (see also: \cite{Zdziarski+2016A&A...595A..52Z}).  For a given eclipse width, there is a degeneracy between the size of the secondary (which varies with q) and the inclination, since a larger secondary can eclipse the \WD for the same duration of time as a smaller would at a higher inclination.  \autoref{fig:iConstraint} illustrates the constraints graphically.  The measured eclipse width and its $1\sigma$~uncertainty is shown spanning a narrow range of possibilities on the y-axis. Also shown are the level contours of the $q(i, \Delta\phi_{\nicefrac{1}{2}})$ relation corresponding to the values in \autoref{tbl:inclination}. We can see from this figure that the $3\sigma$ constraints provided by the Roche geometry are consistent with those provided by the firefly model for $i>82\deg$.  The minimum allowed mass ratio (smallest secondary) corresponding to an inclination of $i=90\deg$ is $q_{min}=0.192$ (the lowest contour in \autoref{fig:iConstraint}).
   
  
  With an inclination of $i > 82 \deg$, we obtain a range of possible mass ratio values ${0.19 \lesssim q \lesssim 0.33}$. With the secondary mass estimated from the evolutionary state, the corresponding range of possible values for the WD mass is ${0.5 \Mo \lesssim M_1 \lesssim 0.86 \Mo}$.  We note that this range is consistent with the distribution of \WD masses in polars (see \cite{2011A&A...536A..42Z, 2015SSRv..191..111F}).  
  
  Using the \cite{Nauenberg1972ApJ...175..417N} \WD mass-radius relation, we obtain the radius of the \WD: ${R_1 = 0.014 \pm 0.002 \Ro}$.  The binary separation is calculated as ${a = 3.25 \pm 0.02 \times 10^{-3}\,}$AU from Kepler's third law.  \autoref{tbl:params} lists the derived system parameters and their uncertainties.

  \begin{table}
    \centering
    \caption{Binary system parameters for \sourceName.}
    \begin{tabular}{*4{l}}
     \hline\rule{0pt}{10pt}%
      $P_{orb}$	&  $2.0786 \pm 0.0009$ h						& $7482.9 \pm 3.5$ s						&		\\
      $M_1$ 	&  $0.5 \Mo \lesssim M_1 \lesssim 0.86 \Mo$		& & \\ 
      $M_2$ 	&  $0.16 \pm 0.03\ \Mo$							& $3.6 \pm 0.6 \times 10^{29}$ kg			&		\\
      $R_1$ 	&  $0.012 \pm 0.002\ \Mo$  						& $1.05 \pm 0.07\times 10^{7}$ m			& 0.021 $a$	\\
      $R_2$ 	&  $0.20 \pm 0.03\ \Ro$							& $1.4 \pm 0.2 \times 10^{8}$ m				& 0.286 $a$ 	\\
      $a$ 		&  $3.25 \pm 0.02\ \times 10^{-3}$ AU			& $4.86 \pm 0.02 \times 10^{8}$ m			& 1 $a$		\\
      $i$		&  $>82 \deg$									&											&		\\
      \hline
     \label{tbl:params}
  	\end{tabular}
  \end{table}

  \section{Discussion}
  \label{sec:discuss}
  
  Polars have been known for some time to show absorption dips in their X-ray light curves. Absorption dips were first noticed in soft X-ray light of EF~Eri \citep{1981ApJ...245..618P}, and have subsequently been observed in the light curves of a number of polars, most notably in V834~Cen, QQ~Vul, AN~UMa \citep{ 1985MNRAS.215P...1K}, QS~Tel \citep{1996MNRAS.280.1121R},  MN~Hya  \citep{1998MNRAS.299..998B}, CE Gru \citep{2002MNRAS.335..918R}, EV~Uma, GG~Leo \citep{2004MNRAS.347...95R} and AI~Tri \citep{2010A&A...516A..76T}.  Recently, \citep{2017RAA....17...10W} also detected absorption dips in the soft and medium X-ray light curve of the long period, low accretion rate polar J215544.4+380116.  A number of IPs also display features in their soft X-ray and extreme-UV light curves characteristic of stream fed accretion near the WD: V709~Cas \citep{2001A&A...377..499D, 2004A&A...415.1009D} and EX~Hya \citep{2002ApJ...577..359B}.

Models that successfully explain the features of X-ray light curves were initially developed in \cite{1985MNRAS.215P...1K} and \cite{1989MNRAS.237..299W} and have subsequently been adapted to explain many features of the EUV light curves as well \citep{1998ApJ...506..824S}. The most consistent idea is that the absorption dips are caused by (partial) occultations of the hot spot and post-shock region by magnetically threaded material at the magnetospheric boundary, or by material in the ballistic portion of the stream. The presence of P~Cygni profiles at phases preceding the eclipse during the high state in HU~Aqr \citep{1997A&A...319..894S} and FL~Cet \citep{2005ApJ...620..422S}  (amongst others) 
further support this interpretation.

Absorption dips in the visible light of Polars, seems to be somewhat less common.  Besides the  example of HU~Aqr, already mentioned in \autoref{sec:lc}, another eclipsing polar, UZ~For, has a pronounced stream absorption dip in the $R$-band light curve and, to a lesser extent, also in the $V$-band \cite{1991MNRAS.253...27B}. The asynchronous polar V1432~Aql has an absorption feature that appears and disappears periodically, depending on the phase of the spin-orbit cycle. \citep{1995PASP..107..307P}.   In general, the shape of the absorption features are strongly wavelength dependent, as can be seen from simultaneous multi-band photometry in \cite{Harrop-Allin1999}. Other recent examples include LSQ1725-64 \cite{2016MNRAS.462.2382F}, a deeply eclipsing polar with a symmetric stream absorption feature appearing during its bright state, as well as CSS081231:071126+440405 \cite{2015AN....336..115S}, another short-period polar with peculiar accretion geometry.

The flow structure in the inner regions of the binary can be quite complex.  The interaction of the accretion stream with the magnetic field can manifest a wide variety of behaviours which depend sensitively on the mass transfer rate, magnetic field structure (strength and direction of the magnetic axis), the orbital period of the system, as well as the nature of the secondary star.  Accretion flows which switch between the one- and two-pole regime has been observed in both QS~Tel \citep{1996MNRAS.280.1121R} and MT~Dra \citep{2002A&A...392..505S}.  Subsequent studies involving magneto-hydrodynamical simulations, have shown that accretion onto quadrupole \cite{2012ARep...56..257Z, 2016AIPC.1714b0002Z} and octupole \cite{2012NewA...17..232L} field topologies are able to produce flows with multiple polar impact zones, or even azimuthally extended hot spots near the WD equator.  There is also strong evidence from Zeeman-tomography that the WDs in (at least two) Polars are able to generate complex multi-polar field structures \citep{Beuermann2007}. 


The way in which the material in the stream interacts with the magnetosphere of the accretor can therefore tell us something about the magnetic field of the primary star, as well as the mass transfer rate and properties of the plasma in the accretion stream.  \cite{2003MNRAS.338..219R}, for example, use the stream absorption dip in the polar RX~J1007.5$-$2017 to constrain the radius and total column density of the stream.  In eclipsing systems in particular, much information is to be gained from the way in which luminous material is systematically mapped out by the occulting limb of the secondary.  Observations of the low state light curves are particularly useful for modelling purposes since discerning the various luminous components in the binary becomes easier as the brightness of the stream and hot spot are diminished relative to the photospheric brightnesses of the stars.  The diminished presence of the post shock accretion column can be seen in the low state light curve of \sourceName in \autoref{fig:folded} by the decrease in brightness across most of the orbit, as well as in the shallower eclipse profile.

Comparing the high-state (2015) and low-state (2017) light curves in \autoref{fig:folded}, we notice a number of additional differences: 
  i) The absorption dip in the low state is narrower by ${\sim}10\deg$ in phase as compared to the bright state. The position of the obscuring part of the stream has therefore shifted or narrowed by $10\deg$ \wrt the line of sight to the bright spot.  This could be explained by a change in the location (radius and/or azimuthal angle) of the threading region, or by a change in the location of the bright spot, or possibly a combination of both.  Under a higher mass transfer regime, we expect the stream to penetrate deeper into the magnetosphere of the WD, and material to be threaded along a wider arc. This would lead to a larger accretion footprint extending to lower magnetic latitudes, which would show up as a phases shift in the position of the dip. We estimate that, at the location of the threading region indicated by the ``firefly'' model, a distance of $10\,R_{WD}$ from the WD, a change of $10\deg$ in azimuth corresponds to a projected distance on the WD surface of ${\sim}0.4\,R_{WD}$.  The accretion column would therefore have had to move a significant distance on the WD surface to reproduce the observed phase shift given an unmoving threading region.  The possibility remains that observed shift is due to a slight asynchronism between the WD spin and the orbital period. In the near-synchronous eclipsing polar, V1432~Aql \citep{Littlefield2015MNRAS.449.3107L}, discrepancies in the eclipse timings have been attributed to changes in the location of the threading region along the ballistic portion of the stream, and it is feasible that the same mechanism may be at work in \sourceName. The absence of any signature of periodic flux modulation due to the WD spin in \autoref{fig:periodogram}, along with the size of the nightly dip-phase shifts during 2015, however, suggest that the changes in the location of the threading region are due to mass transfer variability, rather than secular changes due to spin-orbit asynchronism.
  
  ii) The  shape of the absorption dip remains roughly unchanged between the high- and low-states and is highly asymmetric.  If we assume that the luminosity of the bright spot/post-shock region remain constant across the phase range of the absorption dip, the asymmetry of this feature indicates that the leading edge of the threading region contains denser material than the trailing edge.  This may be an indication of an inhomogeneous, or ``blobby''  accretion flow suggested by \citet{1982A&A...114L...4K, 1995MNRAS.275....9W}, since denser portions of the flow, being somewhat diamagnetic, would follow a ballistic trajectory deeper into the WD magnetosphere and eventually become threaded at a higher magnetic field value.

 iii) The portion of the stream visible after the bright spot has been eclipsed eclipse remains equally bright in both states indicating that there is a significant amount mass transfer even during the low state.

 iv) The average high-state light curve contains a centered at phase~${\sim}0.4$.  The phase of this feature is such that it is located diametrically opposite the location of the primary ingress \ie opposite the bright spot. This feature may be due to an azimuthally extended accretion hot spot visible on the far side of the WD, or potentially to the presence of a second accretion pole. 



    \begin{figure*}
    \centering
    \includegraphics[width=1\textwidth]{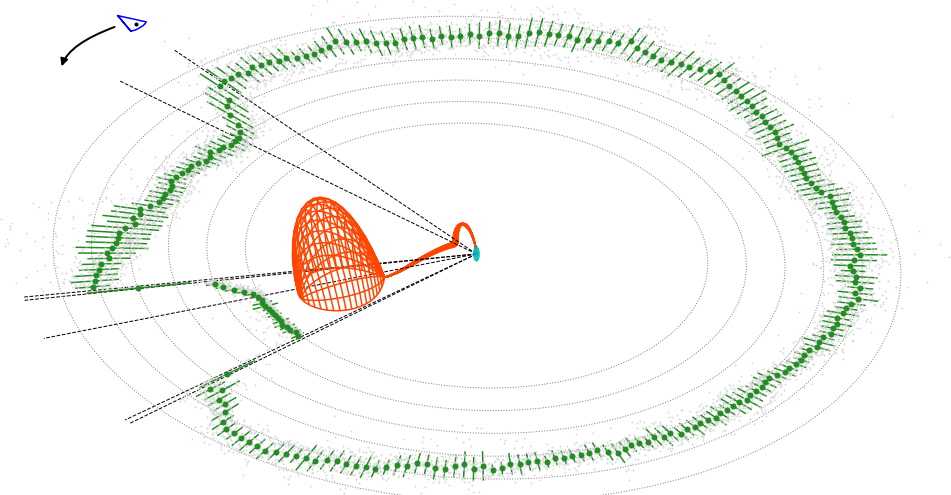} 
    \caption{A three dimensional view into the system geometry. Central in the figure are shown the white dwarf (cyan), the donor star (orange wireframe) and the accretion stream.  A radial projection of the average phase-folded light curve of \sourceName is plotted in green. The phase regions corresponding to those marked in the bottom panel of \autoref{fig:eclipses} (also \autoref{tbl:eclipseParams}) are delineated by the dashed black lines emanating from the \WD along the line of sight to the observer. In the co-rotating reference frame of the binary, the observer would move in a counter-clockwise direction around the center of mass. This figure shows how the features in the light curve correspond to the physical components of the system.}
    \label{fig:3d}
  \end{figure*}

  \section{Conclusions}
  In this paper we have reported on the discovery, observations, and modelling of a new eclipsing \glsentrylong{mCV}, \sourceNameLong. By modelling the light curve with a modified genetic algorithm based on the ``firefly'' model of \citet{Hakala2002MNRAS.334..990H}, we constrain the inclination and mass ratio of the system and subsequently derive the binary orbital parameters.  
  
  In the context of stellar evolution, eclipsing polars such as \sourceName play a particularly important role.  There is by now significant evidence that the evolutionary pathway for mCVs is distinct from that of non-magnetic CVs, perhaps the most conspicuous being the reduced importance of the $2-3\,$h period gap (see \autoref{fig:histogram}).  The period of \sourceName nearly coincides with the lower edge of the period gap - a time in the evolution when the donor star is re-establishing contact with its Roche lobe after being in a prolonged (multi-million year) low-mass-transfer state. The system parameters suggest that \sourceName is most probably evolving to shorter periods, rather than being a post-bounce system, and could therefore be valuable in helping constrain evolutionary scenarios, particularly \wrt the efficacy of residual magnetic breaking within the period gap.  We therefore conclude by recommending \sourceName as a worthy candidate for further multi-wavelength (particularly X-ray) follow up observations in order to shed light on the remaining uncertainties concerning the accretion flow, magnetic field structure, and evolutionary status of \sourceNameLong.
  

  \section*{Acknowledgements}
  Some observations reported in this paper were obtained with the \glsentrylong{SALT} under programme \mbox{2014-2-DDT-002}.  
  DB, HB, MM, AK, SBP and PW acknowledges support through the National Research Foundation of South Africa. AK acknowledges the Russian Science Foundation (project no. 14-50-00043).
  VL and EG acknowledges support of the Development Program of M.V.~Lomonosov Moscow State University and the Russian Science Foundation 16-12-00085. MASTER-SAAO is supported by the Russian Foundation for Basic Research (RFBR)~17-52-80133.  
  We thank Prof.~Jules Halpern of Columbia University, who kindly obtained a short time series in 2017 March from MDM. We also thank the students of the 2015 Dartmouth Astronomy Foreign Study Program for assistance at the telescope; they are MacKenzie Carlson, Z. Edrei Chua, Robert Cueva, Natalie Drozdoff, John French, Emma Garcia, Zoe Guttendorf, Rachel McKee, Krystyna Miles, Jack Neustadt, Samuel Rosen, Marie Schwalbe, and Nicholas Suntzeff, and graduate assistants Erek Alper and MacKenzie Jones.  Finally, we thank the anonymous referee, whose comments and suggestions have lead to greater understanding, and significantly improved the quality of this work.

  
  
  \bibliographystyle{mnras}
  \bibliography{references} 

\begin{thebibliography}{}
\makeatletter
\relax
\def\mn@urlcharsother{\let\do\@makeother \do\$\do\&\do\#\do\^\do\_\do\%\do\~}
\def\mn@doi{\begingroup\mn@urlcharsother \@ifnextchar [ {\mn@doi@}
  {\mn@doi@[]}}
\def\mn@doi@[#1]#2{\def\@tempa{#1}\ifx\@tempa\@empty \href
  {http://dx.doi.org/#2} {doi:#2}\else \href {http://dx.doi.org/#2} {#1}\fi
  \endgroup}
\def\mn@eprint#1#2{\mn@eprint@#1:#2::\@nil}
\def\mn@eprint@arXiv#1{\href {http://arxiv.org/abs/#1} {{\tt arXiv:#1}}}
\def\mn@eprint@dblp#1{\href {http://dblp.uni-trier.de/rec/bibtex/#1.xml}
  {dblp:#1}}
\def\mn@eprint@#1:#2:#3:#4\@nil{\def\@tempa {#1}\def\@tempb {#2}\def\@tempc
  {#3}\ifx \@tempc \@empty \let \@tempc \@tempb \let \@tempb \@tempa \fi \ifx
  \@tempb \@empty \def\@tempb {arXiv}\fi \@ifundefined
  {mn@eprint@\@tempb}{\@tempb:\@tempc}{\expandafter \expandafter \csname
  mn@eprint@\@tempb\endcsname \expandafter{\@tempc}}}

\bibitem[\protect\citeauthoryear{{Bailey} \& {Cropper}}{{Bailey} \&
  {Cropper}}{1991}]{1991MNRAS.253...27B}
{Bailey} J.,  {Cropper} M.,  1991, \mn@doi [\mnras] {10.1093/mnras/253.1.27},
  \href {https://ui.adsabs.harvard.edu/#abs/1991MNRAS.253...27B} {253, 27}

\bibitem[\protect\citeauthoryear{{Beardmore}, {Ramsay}, {Osborne}, {Mason},
  {Nousek}  \& {Baluta}}{{Beardmore} et~al.}{1995}]{1995MNRAS.273..742B}
{Beardmore} A.~P.,  {Ramsay} G.,  {Osborne} J.~P.,  {Mason} K.~O.,  {Nousek}
  J.~A.,   {Baluta} C.,  1995, \mn@doi [\mnras] {10.1093/mnras/273.3.742},
  \href {https://ui.adsabs.harvard.edu/\#abs/1995MNRAS.273..742B} {273, 742}

\bibitem[\protect\citeauthoryear{{Belle}, {Howell}, {Sirk}  \& {Huber}}{{Belle}
  et~al.}{2002}]{2002ApJ...577..359B}
{Belle} K.~E.,  {Howell} S.~B.,  {Sirk} M.~M.,   {Huber} M.~E.,  2002, \mn@doi
  [\apj] {10.1086/342140}, \href
  {https://ui.adsabs.harvard.edu/#abs/2002ApJ...577..359B} {577, 359}

\bibitem[\protect\citeauthoryear{{Beuermann}}{{Beuermann}}{2004}]{Beuermann04}
{Beuermann} K.,  2004, in International Astronomical Union Colloquium.
  Cambridge University Press (\mn@eprint {} {astro-ph/0302235}),
  \mn@doi{10.1017/S0252921100002098}

\bibitem[\protect\citeauthoryear{{Beuermann}, {Euchner}, {Reinsch}, {Jordan}
  \& {G{\"a}nsicke}}{{Beuermann} et~al.}{2007}]{Beuermann2007}
{Beuermann} K.,  {Euchner} F.,  {Reinsch} K.,  {Jordan} S.,   {G{\"a}nsicke}
  B.~T.,  2007, \mn@doi [\aap] {10.1051/0004-6361:20066332}, \href
  {http://adsabs.harvard.edu/abs/2007A%26A...463..647B} {463, 647}

\bibitem[\protect\citeauthoryear{Bradley et~al.,}{Bradley
  et~al.}{2016}]{photutilsv3.0}
Bradley L.,  et~al., 2016, astropy/photutils: v0.3,
  \mn@doi{10.5281/zenodo.164986}, \url {https://doi.org/10.5281/zenodo.164986}

\bibitem[\protect\citeauthoryear{{Bridge}, {Cropper}, {Ramsay}, {de Bruijne},
  {Reynolds}  \& {Perryman}}{{Bridge} et~al.}{2003}]{2003MNRAS.341..863B}
{Bridge} C.~M.,  {Cropper} M.,  {Ramsay} G.,  {de Bruijne} J.~H.~J.,
  {Reynolds} A.~P.,   {Perryman} M.~A.~C.,  2003, \mn@doi [\mnras]
  {10.1046/j.1365-8711.2003.06456.x}, \href
  {https://ui.adsabs.harvard.edu/#abs/2003MNRAS.341..863B} {341, 863}

\bibitem[\protect\citeauthoryear{{Bridge}, {Hakala}, {Cropper}  \&
  {Ramsay}}{{Bridge} et~al.}{2004}]{2004MNRAS.351.1423B}
{Bridge} C.~M.,  {Hakala} P.,  {Cropper} M.,   {Ramsay} G.,  2004, \mn@doi
  [\mnras] {10.1111/j.1365-2966.2004.07884.x}, \href
  {https://ui.adsabs.harvard.edu/#abs/2004MNRAS.351.1423B} {351, 1423}

\bibitem[\protect\citeauthoryear{{Buckley}, {Barrett}, {Haberl}  \&
  {Sekiguchi}}{{Buckley} et~al.}{1998}]{1998MNRAS.299..998B}
{Buckley} D. A.~H.,  {Barrett} P.~E.,  {Haberl} F.,   {Sekiguchi} K.,  1998,
  \mn@doi [\mnras] {10.1046/j.1365-8711.1998.01851.x}, \href
  {https://ui.adsabs.harvard.edu/#abs/1998MNRAS.299..998B} {299, 998}

\bibitem[\protect\citeauthoryear{{Buckley}, {Swart}  \& {Meiring}}{{Buckley}
  et~al.}{2006}]{Buckley2006SPIE.6267E..0ZB}
{Buckley} D.~A.~H.,  {Swart} G.~P.,   {Meiring} J.~G.,  2006, in Society of
  Photo-Optical Instrumentation Engineers (SPIE) Conference Series. p. 62670Z,
  \mn@doi{10.1117/12.673750}

\bibitem[\protect\citeauthoryear{{Buckley} et~al.,}{{Buckley}
  et~al.}{2008}]{Buckley2008SPIE.7014E..07B}
{Buckley} D.~A.~H.,  et~al., 2008, in Ground-based and Airborne Instrumentation
  for Astronomy II. p. 701407, \mn@doi{10.1117/12.789438}

\bibitem[\protect\citeauthoryear{{Burgh}, {Nordsieck}, {Kobulnicky},
  {Williams}, {O'Donoghue}, {Smith}  \& {Percival}}{{Burgh}
  et~al.}{2003}]{Burgh2003SPIE.4841.1463B}
{Burgh} E.~B.,  {Nordsieck} K.~H.,  {Kobulnicky} H.~A.,  {Williams} T.~B.,
  {O'Donoghue} D.,  {Smith} M.~P.,   {Percival} J.~W.,  2003, in {Iye} M.,
  {Moorwood} A.~F.~M.,  eds,  \procspie Vol. 4841, Instrument Design and
  Performance for Optical/Infrared Ground-based Telescopes. pp 1463--1471,
  \mn@doi{10.1117/12.460312}

\bibitem[\protect\citeauthoryear{{Chanan}, {Middleditch}  \& {Nelson}}{{Chanan}
  et~al.}{1976}]{Chanan+1976ApJ...208..512C}
{Chanan} G.~A.,  {Middleditch} J.,   {Nelson} J.~E.,  1976, \mn@doi [\apj]
  {10.1086/154633}, \href
  {https://ui.adsabs.harvard.edu/#abs/1976ApJ...208..512C} {208, 512}

\bibitem[\protect\citeauthoryear{{Coppejans} et~al.,}{{Coppejans}
  et~al.}{2013}]{Coppejans2013PASP..125..976C}
{Coppejans} R.,  et~al., 2013, \mn@doi [\pasp] {10.1086/672156}, \href
  {http://adsabs.harvard.edu/abs/2013PASP..125..976C} {125, 976}

\bibitem[\protect\citeauthoryear{{Crawford} et~al.,}{{Crawford}
  et~al.}{2010}]{Crawford2010SPIE.7737E..25C}
{Crawford} S.~M.,  et~al., 2010, in Observatory Operations: Strategies,
  Processes, and Systems III. p. 773725, \mn@doi{10.1117/12.857000}

\bibitem[\protect\citeauthoryear{{Cropper}}{{Cropper}}{1990}]{Cropper1990}
{Cropper} M.,  1990, \mn@doi [\ssr] {10.1007/BF00177799}, \href
  {http://adsabs.harvard.edu/abs/1990SSRv...54..195C} {54, 195}

\bibitem[\protect\citeauthoryear{{Denisenko} \& {Martinelli}}{{Denisenko} \&
  {Martinelli}}{2016}]{2016arXiv160908511D}
{Denisenko} D.~V.,  {Martinelli} F.,  2016, arXiv e-prints, \href
  {https://ui.adsabs.harvard.edu/\#abs/2016arXiv160908511D} {p.
  arXiv:1609.08511}

\bibitem[\protect\citeauthoryear{{Drake} et~al.,}{{Drake}
  et~al.}{2009}]{Drake2009}
{Drake} A.~J.,  et~al., 2009, \mn@doi [\apj] {10.1088/0004-637X/696/1/870},
  \href {http://adsabs.harvard.edu/abs/2009ApJ...696..870D} {696, 870}

\bibitem[\protect\citeauthoryear{{Drake} et~al.,}{{Drake}
  et~al.}{2014}]{Drake2014}
{Drake} A.~J.,  et~al., 2014, \mn@doi [\apjs] {10.1088/0067-0049/213/1/9},
  \href {http://adsabs.harvard.edu/abs/2014ApJS..213....9D} {213, 9}

\bibitem[\protect\citeauthoryear{{Ferrario}, {de Martino}  \&
  {G{\"a}nsicke}}{{Ferrario} et~al.}{2015a}]{2015SSRv..191..111F}
{Ferrario} L.,  {de Martino} D.,   {G{\"a}nsicke} B.~T.,  2015a, \mn@doi [\ssr]
  {10.1007/s11214-015-0152-0}, \href
  {https://ui.adsabs.harvard.edu/#abs/2015SSRv..191..111F} {191, 111}

\bibitem[\protect\citeauthoryear{Ferrario, de Martino  \&
  G{\"{a}}nsicke}{Ferrario et~al.}{2015b}]{Ferrario2015}
Ferrario L.,  de Martino D.,   G{\"{a}}nsicke B.~T.,  2015b, \mn@doi [Space
  Science Reviews] {10.1007/s11214-015-0152-0}, 191, 111

\bibitem[\protect\citeauthoryear{{Fuchs} et~al.,}{{Fuchs}
  et~al.}{2016}]{2016MNRAS.462.2382F}
{Fuchs} J.~T.,  et~al., 2016, \mn@doi [\mnras] {10.1093/mnras/stw1759}, \href
  {https://ui.adsabs.harvard.edu/#abs/2016MNRAS.462.2382F} {462, 2382}

\bibitem[\protect\citeauthoryear{{Hakala}}{{Hakala}}{1995}]{1995A&A...296..164H}
{Hakala} P.~J.,  1995, \aap, \href
  {http://adsabs.harvard.edu/abs/1995A%26A...296..164H} {296, 164}

\bibitem[\protect\citeauthoryear{{Hakala}, {Piirola}, {Vilhu}, {Osborne}  \&
  {Hannikainen}}{{Hakala} et~al.}{1994}]{Hakala1994MNRAS.271L..41H}
{Hakala} P.~J.,  {Piirola} V.,  {Vilhu} O.,  {Osborne} J.~P.,   {Hannikainen}
  D.~C.,  1994, \mn@doi [\mnras] {10.1093/mnras/271.1.L41}, \href
  {http://adsabs.harvard.edu/abs/1994MNRAS.271L..41H} {271, L41}

\bibitem[\protect\citeauthoryear{{Hakala}, {Cropper}  \& {Ramsay}}{{Hakala}
  et~al.}{2002}]{Hakala2002MNRAS.334..990H}
{Hakala} P.,  {Cropper} M.,   {Ramsay} G.,  2002, \mn@doi [\mnras]
  {10.1046/j.1365-8711.2002.05587.x}, \href
  {http://adsabs.harvard.edu/abs/2002MNRAS.334..990H} {334, 990}

\bibitem[\protect\citeauthoryear{{Harrop-Allin}, {Cropper}, {Hakala}, {Hellier}
   \& {Ramseyer}}{{Harrop-Allin} et~al.}{1999}]{Harrop-Allin1999}
{Harrop-Allin} M.~K.,  {Cropper} M.,  {Hakala} P.~J.,  {Hellier} C.,
  {Ramseyer} T.,  1999, \mn@doi [\mnras] {10.1046/j.1365-8711.1999.02780.x},
  \href {http://adsabs.harvard.edu/abs/1999MNRAS.308..807H} {308, 807}

\bibitem[\protect\citeauthoryear{{Hellier}}{{Hellier}}{2001}]{Hellier2001}
{Hellier} C.,  2001, {Cataclysmic Variable Stars}.
Springer Science \& Business Media

\bibitem[\protect\citeauthoryear{{Hong}, {van den Berg}, {Grindlay},
  {Servillat}  \& {Zhao}}{{Hong} et~al.}{2012}]{2012ApJ...746..165H}
{Hong} J.,  {van den Berg} M.,  {Grindlay} J.~E.,  {Servillat} M.,   {Zhao} P.,
   2012, \mn@doi [\apj] {10.1088/0004-637X/746/2/165}, \href
  {http://adsabs.harvard.edu/abs/2012ApJ...746..165H} {746, 165}

\bibitem[\protect\citeauthoryear{{King} \& {Williams}}{{King} \&
  {Williams}}{1985}]{1985MNRAS.215P...1K}
{King} A.~R.,  {Williams} G.~A.,  1985, \mn@doi [\mnras]
  {10.1093/mnras/215.1.1P}, \href
  {https://ui.adsabs.harvard.edu/#abs/1985MNRAS.215P...1K} {215, 1P}

\bibitem[\protect\citeauthoryear{{Kniazev}, {Pustilnik}, {Grebel}, {Lee}  \&
  {Pramskij}}{{Kniazev} et~al.}{2004}]{Kniazev2004ApJS..153..429K}
{Kniazev} A.~Y.,  {Pustilnik} S.~A.,  {Grebel} E.~K.,  {Lee} H.,   {Pramskij}
  A.~G.,  2004, \mn@doi [\apjs] {10.1086/421519}, \href
  {http://adsabs.harvard.edu/abs/2004ApJS..153..429K} {153, 429}

\bibitem[\protect\citeauthoryear{{Kniazev} et~al.,}{{Kniazev}
  et~al.}{2008}]{Kniazev2008MNRAS.388.1667K}
{Kniazev} A.~Y.,  et~al., 2008, \mn@doi [\mnras]
  {10.1111/j.1365-2966.2008.13435.x}, \href
  {http://adsabs.harvard.edu/abs/2008MNRAS.388.1667K} {388, 1667}

\bibitem[\protect\citeauthoryear{{Knigge}, {Baraffe}  \& {Patterson}}{{Knigge}
  et~al.}{2011}]{Knigge2011ApJS..194...28K}
{Knigge} C.,  {Baraffe} I.,   {Patterson} J.,  2011, \mn@doi [\apjs]
  {10.1088/0067-0049/194/2/28}, \href
  {http://adsabs.harvard.edu/abs/2011ApJS..194...28K} {194, 28}

\bibitem[\protect\citeauthoryear{Kohonen}{Kohonen}{1990}]{Kohonen1990}
Kohonen T.,  1990, \mn@doi [Proceedings of the IEEE] {10.1109/5.58325}, 78,
  1464

\bibitem[\protect\citeauthoryear{{Kornilov} et~al.,}{{Kornilov}
  et~al.}{2012}]{Kornilov2012ExA....33..173K}
{Kornilov} V.~G.,  et~al., 2012, \mn@doi [Experimental Astronomy]
  {10.1007/s10686-011-9280-z}, \href
  {http://adsabs.harvard.edu/abs/2012ExA....33..173K} {33, 173}

\bibitem[\protect\citeauthoryear{{Kuijpers} \& {Pringle}}{{Kuijpers} \&
  {Pringle}}{1982}]{1982A&A...114L...4K}
{Kuijpers} J.,  {Pringle} J.~E.,  1982, \aap, \href
  {https://ui.adsabs.harvard.edu/#abs/1982A&A...114L...4K} {114, L4}

\bibitem[\protect\citeauthoryear{{Lipunov} et~al.,}{{Lipunov}
  et~al.}{2010}]{Lipunov2010AdAst2010E..30L}
{Lipunov} V.,  et~al., 2010, \mn@doi [Advances in Astronomy]
  {10.1155/2010/349171}, \href
  {http://adsabs.harvard.edu/abs/2010AdAst2010E..30L} {2010, 349171}

\bibitem[\protect\citeauthoryear{{Lipunov} et~al.,}{{Lipunov}
  et~al.}{2016}]{Lipunov2016RMxAC..48...42L}
{Lipunov} V.,  et~al., 2016, in Revista Mexicana de Astronomia y Astrofisica
  Conference Series. pp 42--47

\bibitem[\protect\citeauthoryear{{Littlefield} et~al.,}{{Littlefield}
  et~al.}{2015}]{Littlefield2015MNRAS.449.3107L}
{Littlefield} C.,  et~al., 2015, \mn@doi [\mnras] {10.1093/mnras/stv462}, \href
  {http://cdsads.u-strasbg.fr/abs/2015MNRAS.449.3107L} {449, 3107}

\bibitem[\protect\citeauthoryear{{Lomb}}{{Lomb}}{1976}]{1976Ap&SS..39..447L}
{Lomb} N.~R.,  1976, \mn@doi [\apss] {10.1007/BF00648343}, \href
  {http://cdsads.u-strasbg.fr/abs/1976Ap%26SS..39..447L} {39, 447}

\bibitem[\protect\citeauthoryear{{Long}, {Romanova}  \& {Lamb}}{{Long}
  et~al.}{2012}]{2012NewA...17..232L}
{Long} M.,  {Romanova} M.~M.,   {Lamb} F.~K.,  2012, \mn@doi [\na]
  {10.1016/j.newast.2011.08.001}, \href
  {https://ui.adsabs.harvard.edu/#abs/2012NewA...17..232L} {17, 232}

\bibitem[\protect\citeauthoryear{{Masetti} et~al.,}{{Masetti}
  et~al.}{2013}]{Masetti2013A&A...556A.120M}
{Masetti} N.,  et~al., 2013, \mn@doi [\aap] {10.1051/0004-6361/201322026},
  \href {http://adsabs.harvard.edu/abs/2013A%26A...556A.120M} {556, A120}

\bibitem[\protect\citeauthoryear{{Mason}}{{Mason}}{2004}]{Mason2004RMxAC..20..180M}
{Mason} P.~A.,  2004, in {Tovmassian} G.,  {Sion} E.,  eds,  Revista Mexicana
  de Astronomia y Astrofisica, vol.~27 Vol. 20, Revista Mexicana de Astronomia
  y Astrofisica Conference Series. pp 180--181

\bibitem[\protect\citeauthoryear{{Mason}, {Ramsay}, {Andronov}, {Kolesnikov},
  {Shakhovskoy}  \& {Pavlenko}}{{Mason} et~al.}{1998}]{1998MNRAS.295..511M}
{Mason} P.~A.,  {Ramsay} G.,  {Andronov} I.,  {Kolesnikov} S.,  {Shakhovskoy}
  N.,   {Pavlenko} E.,  1998, \mn@doi [\mnras]
  {10.1046/j.1365-8711.1998.01185.x}, \href
  {https://ui.adsabs.harvard.edu/\#abs/1998MNRAS.295..511M} {295, 511}

\bibitem[\protect\citeauthoryear{McAllister et~al.,}{McAllister
  et~al.}{2017}]{McAllister+2017}
McAllister M.~J.,  et~al., 2017, \mn@doi [Monthly Notices of the Royal
  Astronomical Society] {10.1093/mnras/stw2417}, 464, 1353

\bibitem[\protect\citeauthoryear{{Mouchet} et~al.,}{{Mouchet}
  et~al.}{2017}]{2017A&A...600A..53M}
{Mouchet} M.,  et~al., 2017, \mn@doi [\aap] {10.1051/0004-6361/201630166},
  \href {http://adsabs.harvard.edu/abs/2017A%26A...600A..53M} {600, A53}

\bibitem[\protect\citeauthoryear{{Mukai}}{{Mukai}}{2017}]{Mukai2017PASP..129f2001M}
{Mukai} K.,  2017, \mn@doi [\pasp] {10.1088/1538-3873/aa6736}, \href
  {http://cdsads.u-strasbg.fr/abs/2017PASP..129f2001M} {129, 062001}

\bibitem[\protect\citeauthoryear{{Nauenberg}}{{Nauenberg}}{1972}]{Nauenberg1972ApJ...175..417N}
{Nauenberg} M.,  1972, \mn@doi [\apj] {10.1086/151568}, \href
  {http://adsabs.harvard.edu/abs/1972ApJ...175..417N} {175, 417}

\bibitem[\protect\citeauthoryear{{O'Donoghue} et~al.,}{{O'Donoghue}
  et~al.}{2006}]{O'Donoghue2006}
{O'Donoghue} D.,  et~al., 2006, \mn@doi [\mnras]
  {10.1111/j.1365-2966.2006.10834.x}, \href
  {http://adsabs.harvard.edu/abs/2006MNRAS.372..151O} {372, 151}

\bibitem[\protect\citeauthoryear{{Patterson}}{{Patterson}}{1994}]{Patterson1994PASP..106..209P}
{Patterson} J.,  1994, \mn@doi [\pasp] {10.1086/133375}, \href
  {http://adsabs.harvard.edu/abs/1994PASP..106..209P} {106, 209}

\bibitem[\protect\citeauthoryear{{Patterson}, {Williams}  \&
  {Hiltner}}{{Patterson} et~al.}{1981}]{1981ApJ...245..618P}
{Patterson} J.,  {Williams} G.,   {Hiltner} W.~A.,  1981, \mn@doi [\apj]
  {10.1086/158837}, \href
  {https://ui.adsabs.harvard.edu/#abs/1981ApJ...245..618P} {245, 618}

\bibitem[\protect\citeauthoryear{{Patterson}, {Skillman}, {Thorstensen}  \&
  {Hellier}}{{Patterson} et~al.}{1995}]{1995PASP..107..307P}
{Patterson} J.,  {Skillman} D.~R.,  {Thorstensen} J.,   {Hellier} C.,  1995,
  \mn@doi [Publications of the Astronomical Society of the Pacific]
  {10.1086/133556}, \href
  {https://ui.adsabs.harvard.edu/#abs/1995PASP..107..307P} {107, 307}

\bibitem[\protect\citeauthoryear{{Ramsay} \& {Cropper}}{{Ramsay} \&
  {Cropper}}{2002}]{2002MNRAS.335..918R}
{Ramsay} G.,  {Cropper} M.,  2002, \mn@doi [\mnras]
  {10.1046/j.1365-8711.2002.05666.x}, \href
  {https://ui.adsabs.harvard.edu/#abs/2002MNRAS.335..918R} {335, 918}

\bibitem[\protect\citeauthoryear{{Ramsay} \& {Cropper}}{{Ramsay} \&
  {Cropper}}{2003}]{2003MNRAS.338..219R}
{Ramsay} G.,  {Cropper} M.,  2003, \mn@doi [\mnras]
  {10.1046/j.1365-8711.2003.06045.x}, \href
  {https://ui.adsabs.harvard.edu/#abs/2003MNRAS.338..219R} {338, 219}

\bibitem[\protect\citeauthoryear{{Ramsay}, {Cropper}, {Mason}, {C{\'o}rdova}
  \& {Priedhorsky}}{{Ramsay} et~al.}{2004}]{2004MNRAS.347...95R}
{Ramsay} G.,  {Cropper} M.,  {Mason} K.~O.,  {C{\'o}rdova} F.~A.,
  {Priedhorsky} W.,  2004, \mn@doi [\mnras] {10.1111/j.1365-2966.2004.07242.x},
  \href {https://ui.adsabs.harvard.edu/#abs/2004MNRAS.347...95R} {347, 95}

\bibitem[\protect\citeauthoryear{{Ritter} \& {Kolb}}{{Ritter} \&
  {Kolb}}{2003}]{Ritter2003}
{Ritter} H.,  {Kolb} U.,  2003, \mn@doi [\aap] {10.1051/0004-6361:20030330},
  \href {http://adsabs.harvard.edu/abs/2003A%26A...404..301R} {404, 301}

\bibitem[\protect\citeauthoryear{{Rojas} et~al.,}{{Rojas}
  et~al.}{2017}]{Rojas2017A&A...602A.124R}
{Rojas} A.~F.,  et~al., 2017, \mn@doi [\aap] {10.1051/0004-6361/201629463},
  \href {http://adsabs.harvard.edu/abs/2017A%26A...602A.124R} {602, A124}

\bibitem[\protect\citeauthoryear{{Rosen} et~al.,}{{Rosen}
  et~al.}{1996}]{1996MNRAS.280.1121R}
{Rosen} S.~R.,  et~al., 1996, \mn@doi [\mnras] {10.1093/mnras/280.4.14.1},
  \href {https://ui.adsabs.harvard.edu/#abs/1996MNRAS.280.1121R} {280, 1121}

\bibitem[\protect\citeauthoryear{Rosner}{Rosner}{1983}]{Rosner83}
Rosner B.,  1983, \mn@doi [Technometrics] {10.1080/00401706.1983.10487848}, 25,
  165

\bibitem[\protect\citeauthoryear{{Salvi}, {Ramsay}, {Cropper}, {Buckley}  \&
  {Stobie}}{{Salvi} et~al.}{2002}]{2002MNRAS.331..488S}
{Salvi} N.,  {Ramsay} G.,  {Cropper} M.,  {Buckley} D.~A.~H.,   {Stobie} R.~S.,
   2002, \mn@doi [\mnras] {10.1046/j.1365-8711.2002.05216.x}, \href
  {https://ui.adsabs.harvard.edu/\#abs/2002MNRAS.331..488S} {331, 488}

\bibitem[\protect\citeauthoryear{{Scargle}}{{Scargle}}{1982}]{1982ApJ...263..835S}
{Scargle} J.~D.,  1982, \mn@doi [\apj] {10.1086/160554}, \href
  {http://cdsads.u-strasbg.fr/abs/1982ApJ...263..835S} {263, 835}

\bibitem[\protect\citeauthoryear{{Schmidt}, {Hoard}, {Szkody}, {Melia},
  {Honeycutt}  \& {Wagner}}{{Schmidt} et~al.}{1999}]{1999ApJ...525..407S}
{Schmidt} G.~D.,  {Hoard} D.~W.,  {Szkody} P.,  {Melia} F.,  {Honeycutt} R.~K.,
    {Wagner} R.~M.,  1999, \mn@doi [\apj] {10.1086/307901}, \href
  {https://ui.adsabs.harvard.edu/\#abs/1999ApJ...525..407S} {525, 407}

\bibitem[\protect\citeauthoryear{{Schmidt} et~al.,}{{Schmidt}
  et~al.}{2005}]{2005ApJ...620..422S}
{Schmidt} G.~D.,  et~al., 2005, \mn@doi [\apj] {10.1086/426807}, \href
  {https://ui.adsabs.harvard.edu/#abs/2005ApJ...620..422S} {620, 422}

\bibitem[\protect\citeauthoryear{{Schwarz}, {Greiner}, {Tovmassian}, {Zharikov}
   \& {Wenzel}}{{Schwarz} et~al.}{2002}]{2002A&A...392..505S}
{Schwarz} R.,  {Greiner} J.,  {Tovmassian} G.~H.,  {Zharikov} S.~V.,   {Wenzel}
  W.,  2002, \mn@doi [\aap] {10.1051/0004-6361:20021193}, \href
  {https://ui.adsabs.harvard.edu/#abs/2002A&A...392..505S} {392, 505}

\bibitem[\protect\citeauthoryear{{Schwope}, {Thomas}  \& {Beuermann}}{{Schwope}
  et~al.}{1993}]{1993A&A...271L..25S}
{Schwope} A.~D.,  {Thomas} H.~C.,   {Beuermann} K.,  1993, \aap, \href
  {https://ui.adsabs.harvard.edu/\#abs/1993A&A...271L..25S} {271, L25}

\bibitem[\protect\citeauthoryear{{Schwope}, {Thomas}, {Beuermann}, {Burwitz},
  {Jordan}  \& {Haefner}}{{Schwope} et~al.}{1995}]{1995A&A...293..764S}
{Schwope} A.~D.,  {Thomas} H.-C.,  {Beuermann} K.,  {Burwitz} V.,  {Jordan} S.,
    {Haefner} R.,  1995, \aap, \href
  {http://adsabs.harvard.edu/abs/1995A%26A...293..764S} {293, 764}

\bibitem[\protect\citeauthoryear{{Schwope}, {Mantel}  \& {Horne}}{{Schwope}
  et~al.}{1997}]{1997A&A...319..894S}
{Schwope} A.~D.,  {Mantel} K.~H.,   {Horne} K.,  1997, \aap, \href
  {https://ui.adsabs.harvard.edu/#abs/1997A&A...319..894S} {319, 894}

\bibitem[\protect\citeauthoryear{{Schwope}, {Mackebrandt}, {Thinius},
  {Littlefield}, {Garnavich}, {Oksanen}  \& {Granzer}}{{Schwope}
  et~al.}{2015}]{2015AN....336..115S}
{Schwope} A.~D.,  {Mackebrandt} F.,  {Thinius} B.~D.,  {Littlefield} C.,
  {Garnavich} P.,  {Oksanen} A.,   {Granzer} T.,  2015, \mn@doi [Astronomische
  Nachrichten] {10.1002/asna.201412151}, \href
  {https://ui.adsabs.harvard.edu/#abs/2015AN....336..115S} {336, 115}

\bibitem[\protect\citeauthoryear{{Shumkov} et~al.,}{{Shumkov}
  et~al.}{2015}]{Shumkov2015ATel.7127....1S}
{Shumkov} V.,  et~al., 2015, The Astronomer's Telegram, \href
  {http://adsabs.harvard.edu/abs/2015ATel.7127....1S} {7127}

\bibitem[\protect\citeauthoryear{{Sirk} \& {Howell}}{{Sirk} \&
  {Howell}}{1998}]{1998ApJ...506..824S}
{Sirk} M.~M.,  {Howell} S.~B.,  1998, \mn@doi [\apj] {10.1086/306264}, \href
  {https://ui.adsabs.harvard.edu/#abs/1998ApJ...506..824S} {506, 824}

\bibitem[\protect\citeauthoryear{{Spark} \& {O'Donoghue}}{{Spark} \&
  {O'Donoghue}}{2015}]{Spark+2015MNRAS.449..175S}
{Spark} M.~K.,  {O'Donoghue} D.,  2015, \mn@doi [\mnras]
  {10.1093/mnras/stv233}, \href
  {http://adsabs.harvard.edu/abs/2015MNRAS.449..175S} {449, 175}

\bibitem[\protect\citeauthoryear{Stickel}{Stickel}{2010}]{Stickel2010}
Stickel J.~J.,  2010, \mn@doi [Computers & Chemical Engineering]
  {https://doi.org/10.1016/j.compchemeng.2009.10.007}, 34, 467

\bibitem[\protect\citeauthoryear{{Szkody}}{{Szkody}}{1998}]{Szkody1998ASPC..137...18S}
{Szkody} P.,  1998, in {Howell} S.,  {Kuulkers} E.,   {Woodward} C.,  eds,
  Astronomical Society of the Pacific Conference Series Vol. 137, Wild Stars in
  the Old West. p.~18

\bibitem[\protect\citeauthoryear{{Traulsen}, {Reinsch}, {Schwarz}, {Dreizler},
  {Beuermann}, {Schwope}  \& {Burwitz}}{{Traulsen}
  et~al.}{2010}]{2010A&A...516A..76T}
{Traulsen} I.,  {Reinsch} K.,  {Schwarz} R.,  {Dreizler} S.,  {Beuermann} K.,
  {Schwope} A.~D.,   {Burwitz} V.,  2010, \mn@doi [\aap]
  {10.1051/0004-6361/200913201}, \href
  {https://ui.adsabs.harvard.edu/#abs/2010A&A...516A..76T} {516, A76}

\bibitem[\protect\citeauthoryear{{VanderPlas} \& {Ivezi{\'c}}}{{VanderPlas} \&
  {Ivezi{\'c}}}{2015}]{VanderPlas2015ApJ...812...18V}
{VanderPlas} J.~T.,  {Ivezi{\'c}} {\v Z}.,  2015, \mn@doi [\apj]
  {10.1088/0004-637X/812/1/18}, \href
  {http://adsabs.harvard.edu/abs/2015ApJ...812...18V} {812, 18}

\bibitem[\protect\citeauthoryear{Vanderplas}{Vanderplas}{2015}]{jake_vanderplas_2015_14833}
Vanderplas J.,  2015, {gatspy: General tools for Astronomical Time Series in
  Python}, \mn@doi{10.5281/zenodo.14833}, \url
  {https://doi.org/10.5281/zenodo.14833}

\bibitem[\protect\citeauthoryear{{Wang}, {Bai}, {Zhang}  \& {Liu}}{{Wang}
  et~al.}{2017}]{2017RAA....17...10W}
{Wang} S.,  {Bai} Y.,  {Zhang} C.-P.,   {Liu} J.-F.,  2017, \mn@doi [Research
  in Astronomy and Astrophysics] {10.1088/1674-4527/17/1/10}, \href
  {https://ui.adsabs.harvard.edu/#abs/2017RAA....17...10W} {17, 10}

\bibitem[\protect\citeauthoryear{{Warner}}{{Warner}}{2003}]{Warner2003}
{Warner} B.,  2003, {Cataclysmic Variable Stars}.
Cambridge University Press, \mn@doi{10.1017/CB0978O511586491}

\bibitem[\protect\citeauthoryear{{Watson}, {King}, {Jones}  \&
  {Motch}}{{Watson} et~al.}{1989}]{1989MNRAS.237..299W}
{Watson} M.~G.,  {King} A.~R.,  {Jones} M.~H.,   {Motch} C.,  1989, \mn@doi
  [\mnras] {10.1093/mnras/237.1.299}, \href
  {https://ui.adsabs.harvard.edu/#abs/1989MNRAS.237..299W} {237, 299}

\bibitem[\protect\citeauthoryear{{Wynn} \& {King}}{{Wynn} \&
  {King}}{1995}]{1995MNRAS.275....9W}
{Wynn} G.~A.,  {King} A.~R.,  1995, \mn@doi [\mnras] {10.1093/mnras/275.1.9},
  \href {https://ui.adsabs.harvard.edu/#abs/1995MNRAS.275....9W} {275, 9}

\bibitem[\protect\citeauthoryear{{Zdziarski}, {Zi{\'o}{\l}kowski}, {Bozzo}  \&
  {Pjanka}}{{Zdziarski} et~al.}{2016}]{Zdziarski+2016A&A...595A..52Z}
{Zdziarski} A.~A.,  {Zi{\'o}{\l}kowski} J.,  {Bozzo} E.,   {Pjanka} P.,  2016,
  \mn@doi [\aap] {10.1051/0004-6361/201628585}, \href
  {https://ui.adsabs.harvard.edu/#abs/2016A&A...595A..52Z} {595, A52}

\bibitem[\protect\citeauthoryear{{Zhilkin}, {Bisikalo}  \& {Mason}}{{Zhilkin}
  et~al.}{2012}]{2012ARep...56..257Z}
{Zhilkin} A.~G.,  {Bisikalo} D.~V.,   {Mason} P.~A.,  2012, \mn@doi [Astronomy
  Reports] {10.1134/S1063772912040087}, \href
  {https://ui.adsabs.harvard.edu/#abs/2012ARep...56..257Z} {56, 257}

\bibitem[\protect\citeauthoryear{{Zhilkin}, {Bisikalo}  \& {Mason}}{{Zhilkin}
  et~al.}{2016}]{2016AIPC.1714b0002Z}
{Zhilkin} A.~G.,  {Bisikalo} D.~V.,   {Mason} P.~A.,  2016, in American
  Institute of Physics Conference Series. p. 020002, \mn@doi{10.1063/1.4942564}

\bibitem[\protect\citeauthoryear{{Zorotovic}, {Schreiber}  \&
  {G{\"a}nsicke}}{{Zorotovic} et~al.}{2011}]{2011A&A...536A..42Z}
{Zorotovic} M.,  {Schreiber} M.~R.,   {G{\"a}nsicke} B.~T.,  2011, \mn@doi
  [\aap] {10.1051/0004-6361/201116626}, \href
  {https://ui.adsabs.harvard.edu/#abs/2011A&A...536A..42Z} {536, A42}

\bibitem[\protect\citeauthoryear{{de Martino} et~al.,}{{de Martino}
  et~al.}{2001}]{2001A&A...377..499D}
{de Martino} D.,  et~al., 2001, \mn@doi [\aap] {10.1051/0004-6361:20011059},
  \href {https://ui.adsabs.harvard.edu/#abs/2001A&A...377..499D} {377, 499}

\bibitem[\protect\citeauthoryear{{de Martino}, {Matt}, {Belloni}, {Haberl}  \&
  {Mukai}}{{de Martino} et~al.}{2004}]{2004A&A...415.1009D}
{de Martino} D.,  {Matt} G.,  {Belloni} T.,  {Haberl} F.,   {Mukai} K.,  2004,
  \mn@doi [\aap] {10.1051/0004-6361:20034160}, \href
  {https://ui.adsabs.harvard.edu/#abs/2004A&A...415.1009D} {415, 1009}

\makeatother
\end{thebibliography}

  
  
  
  %
  %
  

  \bsp	
  \label{lastpage}
\end{document}